\documentclass[preprint,12pt,3p]{elsarticle}
\usepackage{amssymb}
\usepackage{amsmath}
\usepackage{graphicx}
\usepackage{dcolumn}
\usepackage{bm}
\usepackage{multicol}
\usepackage[utf8]{inputenc}
\usepackage[T1]{fontenc}
\usepackage{etoolbox}
\usepackage{xcolor}
\usepackage{amsfonts}
\usepackage{stmaryrd}
\usepackage{dsfont}
\usepackage{float}
\usepackage{subfig}
\usepackage{subcaption}
\usepackage{microtype}
\usepackage{changepage}
\usepackage[hidelinks]{hyperref}

\DeclareMathOperator{\Tr}{Tr}

\usepackage{tikz}
\usepackage{pgfplots}
\usepgfplotslibrary{colormaps}
\usepgfplotslibrary{colorbrewer}
\pgfplotsset{cycle list/Dark2}
\pgfplotsset{compat=1.17}

\journal{Computer Physics Communications}

\begin{document}

\begin{frontmatter}

\title{Implementation of Asymptotic Preserving Discrete Velocity Methods into the Simulation Code PICLas}

\author[IRS,1]{Félix Garmirian}
\ead{garmirianf@irs.uni-stuttgart.de}
\author[IRS]{Marcel Pfeiffer}

\affiliation[IRS]{organization={Institute of Space Systems, University of Stuttgart},
            addressline={Pfaffenwaldring 29, 70569 Stuttgart},
            country={Germany}}
\affiliation[1]{Corresponding author}

\begin{abstract}
  \setcitestyle{authoryear}
  The Bhatnagar-Gross-Krook (BGK) model of the Boltzmann equation allows for efficient flow simulations, especially in the transition regime between continuum and high rarefaction. However, ensuring efficient performances for multiscale flows, in which the Knudsen number varies by several orders of magnitude, is never straightforward. Discrete velocity methods as well as particle-based solvers can each reveal advantageous in different conditions, but not without compromises in specific regimes. This article presents a second-order asymptotic preserving discrete velocity method to solve the BGK equation, with the particularity of maintaining positivity when operations are conducted with the cell-local distribution function. With this procedure based on exponential differencing, it is therefore also possible to construct an adapted version of this second-order method using the stochastic particle approach, as presented in~\citet{pfeiffer2022exponential}. The deterministic variant and its implementation are detailed here and its performances are evaluated on several test cases. Combined to the probabilistic solver and with the possibility of a future coupling, our exponential differencing discrete velocity method provides a robust toolbox, useful for efficiently simulating multiscale gas phenomena.
\end{abstract}

\begin{keyword}
Discrete velocity method \sep BGK equation \sep Multiscale numerical method \sep Rarefied gas dynamics


\end{keyword}

\end{frontmatter}

\section{Introduction}
In recent years, a variety of gas kinetic methods allowing for efficient multiscale non-equilibrium simulations have been developed. Different classifications of such methods can be introduced. One possible classification is the division between mostly deterministic methods in Euler formulation, i.e. on a fixed grid~\cite{guo2013discrete,xu2010unified,mieussens2000discrete}, and mostly stochastic methods in Lagrange formulation, i.e. on a moving grid, which are usually also referred to as particle methods~\cite{gallis2011investigation,pfeiffer2018particlebased,fei2020unified,fei2021efficient,zhang2019particlebased,pfeiffer2019evaluation,gorji2014efficient}, with some methods combining both representations~\cite{liu2020unified,yang2023discrete}.

Among the methods with an Euler formulation, the unified gas kinetic scheme~\cite{xu2010unified} (UGKS) and discrete unified gas kinetic scheme~\cite{guo2013discrete,guo2015discrete,guo2021progress} (DUGKS) as special forms of the discrete velocity method~\cite{mieussens2000discrete} (DVM) have recently made great progress in the efficient simulation of multiscale problems. As a significant feature, both methods show an asymptotic preserving (AP) behaviour. The time step size is thus not limited to the relaxation time due to the implicit treatment of the collision operator (here a BGK operator). Besides, for small Knudsen numbers the behaviour of the Chapman-Enskog expansion is preserved and the methods are at least second-order accurate in both the free-molecular regime and the continuum regime. The implicit treatment of the collision term is realized by a Crank-Nicolson integration, which can be written down again in an explicit form by clever variable transformation, resulting in a particularly efficient method~\cite{guo2013discrete,xu2010unified}.

For particle methods on the other hand, there have also been efforts in recent years to construct AP methods. The collision term of the complicated Boltzmann equation is again often replaced with simpler approximations, such as the BGK~\cite{gallis2011investigation,pfeiffer2018particlebased,fei2020unified,fei2021efficient,pfeiffer2022exponential} or the Fokker-Planck operator~\cite{gorji2011fokker,gorji2014efficient, pfeiffer2017adaptive, mathiaud2016fokker}. In the case of the BGK operator in particular, one could try to obtain an AP method by using the approach from the DUGKS method with the Crank-Nicolson integration and transfer it to the particles. However, since the integration would then lead to a multiplication of the distribution function with pre-factors that can take on any (positive or negative) value, as shown in Section~\ref{sec:DUGKS}, this approach is not easily applicable to particle methods.

One solution to this problem is the unified stochastic particle BGK (USP-BGK) method~\cite{fei2020unified,fei2021efficient}. Here, an additional collision term was inserted in which the current distribution function is approximated by a Grad-13 approximation~\cite{struchtrup2003regularization}. This allows the advection and the relaxation process to be solved together, which is a requirement for the construction of AP methods. The choice of the Grad-13 distribution also ensures that the Navier-Stokes limit is asymptotically preserved. This additional collision term is constructed to satisfy the Navier-Stokes equations in the continuum domain with a second-order time integration. In the case of large Knudsen numbers, however, the method falls back to first order.

A similar idea is used in the construction of the Crank-Nicolson stochastic particle BGK method~\cite{pfeiffer2025cranka}, which, for low enough time steps, adapts DUGKS to the particle-based approach. For time steps larger than the relaxation time, the relaxation step is replaced by a sampling from a Grad-13 distribution, built with accordingly evolved higher moments, to avoid the problem of negative weights. This allows to retain the second-order accuracy of DUGKS, even towards the free-molecular limit.

A different possibility is given by the unified gas kinetic wave-particle (UGKWP) method~\cite{liu2020unified} and its simplified version (SUWP)~\cite{liu2022progress,yang2024simplified} where particles are classified at each time step into collisional and free-transport particles, the collisional part being treated hydrodynamically. These "wave-particle" methods have the additional advantage of decreasing the simulation particle number in near-continuum regions of the flow, thereby reducing the statistical noise in the results.

Another approach that leads to a second-order particle method in both rarefied and continuum regimes was recently presented in the form of the exponential integration of the BGK equation~\cite{pfeiffer2022exponential}. With this method, thanks to certain pre-factors as the result of implicit integration, it is very easy to construct particle processes that always have positive particle weights and AP behaviour.

In this article, this exponential integration approach will be used to construct an AP-DVM method similar to DUGKS. The idea of exponential integration has been successfully used in the past to construct asymptotic preserving Runge-Kutta methods of second order~\cite{hu2019second} or even higher order for the BGK equation~\cite{dimarco2011exponential,boscheri2020high, li2014exponential}. The special feature of the integrator presented here is that it has then the exact same integration approach and behavior as the particle method of~\citet{pfeiffer2022exponential}. It should therefore be possible to easily couple the DVM approach and the particle approach in further work, in order to be able to use the advantages of both approaches in the flow simulations. This idea is already used in the discrete UGKWP method~\cite{yang2023discrete} to combine the advantages of DVM and particle methods. Here, however, we would have the first approach in which both methods can be integrated identically and equally in time, hopefully simplifying the coupling and its behaviour.

The devised DVM is therefore implemented, along with DUGKS, in the open-source code PICLas~\cite{fasoulas2019combining} which is already an extensive framework for particle-based simulations and which is available on GitHub\footnote{\url{https://github.com/piclas-framework/piclas}}.
In PICLas, the DVM solver can now be selected as a modular compile flag as part of the existing finite volume framework. In addition, the possibility of coupling the DVM module with other modules such as the Hybridisable Discontinuous Galerkin module for solving the Poisson equation and simulation of plasmas~\cite{pfeiffer2019particleincella} or the various particle modules (Direct Simulation Monte Carlo~\cite{bird1994molecular}, BGK~\cite{pfeiffer2019evaluation,hild2024multispecies}, Particle-In-Cell~\cite{pfeiffer2019particleincella}) is currently being worked on.

This article begins with an overview of the BGK approximation and a summary of the DUGKS method in order to go into the problem of the pre-factors again and derive our new method (Section \ref{sec:theory}). Then, the implementation of the exponential integrator method for the DVM approach is shown (Section \ref{sec:implementation}), followed by various validation cases (Section \ref{sec:results}).

\section{Theory}
\label{sec:theory}

\subsection{BGK approximation}

The basis for the kinetic description of gases is the Boltzmann equation
\begin{equation}
  \frac{\partial f}{\partial t} + \mathbf v \cdot \frac{\partial f}{\partial \mathbf x} = \left.\frac{\partial f}{\partial t}\right|_{\mathrm{coll}},
  \label{eq:Boltzmann}
\end{equation}
given here in the monatomic case, where the distribution function $f(\mathbf x, \mathbf v, t)$ depends on time $t$, particle velocity $\mathbf v$ and position $\mathbf x$.

The BGK model~\cite{bhatnagar1954model} approximates the collision term by making the distribution function $f$ relax towards a target distribution $f^t$, with a certain relaxation frequency $\nu$:
\begin{equation}
      \left.\frac{\partial f}{\partial t}\right|_{\mathrm{coll}} = \nu (f^t - f)=\Omega.
      \label{eq:CollOp}
\end{equation}
A first choice of target would be a simple Maxwellian distribution
\begin{equation}
  f^M=\rho\left(\frac{1}{2\pi R T}\right)^{3/2} \exp\left[-\frac{\mathbf c \cdot \mathbf c}{2R T}\right],
  \label{eq:maxwell}
\end{equation}
with density $\rho$, temperature $T$, specific gas constant $R$ and the thermal velocity $\mathbf c = \mathbf v - \mathbf u$, where $\mathbf u$ is the average flow velocity.
The relaxation frequency is chosen so as to give rise to the intended viscosity $\mu=\frac{\rho R T}{\nu}$. However, this leads to a fixed Prandtl number $\mathrm{Pr}=1$.

To obtain a correct Prandtl number ($2/3$ for a monatomic ideal gas), one option is the Shakhov model~\cite{shakhov1968generalization}. In this model, the heat flux $\mathbf q(f)=\frac{1}{2}\int \mathbf c (\mathbf c\cdot \mathbf c)fd\mathbf v$ is used to modify the target distribution in order to accurately account for viscous and thermal effects at the same time:
\begin{equation}
  f^S=f^M\left[1+(1-\mathrm{Pr})\frac{\mathbf c \cdot \mathbf q}{5\rho(RT)^2}\left(\frac{\mathbf c^2}{RT}-5\right)\right].
\end{equation}

Another possibility is the ellipsoidal statistical BGK (ESBGK) model~\cite{holway1966new} in which the relaxation frequency is modified into $\nu=\frac{\rho RT\mathrm{Pr}}{\mu}$ to retrieve the correct heat flux relaxation rate, while the use of the pressure tensor $\mathcal{P}=\int \mathbf c \mathbf c^T f d\mathbf v$ in the target distribution $f^{ES}$ ensures correct viscous effects:
\begin{equation}
f^{ES}=\frac{\rho}{\sqrt{\det (2\pi\mathcal{A})}}\exp\left[-\frac{1}{2}\mathbf c^T\mathcal A^{-1}\mathbf c\right]
\label{eq:esbgkdist}
\end{equation}
with the matrix
\begin{equation}
\mathcal{A} = \frac{RT}{\mathrm{Pr}}\mathcal{I} + \left(1-\frac{1}{\mathrm{Pr}}\right)\frac{\mathcal P}{\rho}
\end{equation}
where $\mathcal{I}$ is the identity matrix.

\subsection{Recap of the Discrete Unified Gas Kinetic Scheme}
\label{sec:DUGKS}
The idea behind the DUGKS is to construct an asymptotic preserving integration method to solve the BGK equation with second-order accuracy~\cite{guo2013discrete,guo2015discrete}. For this, the velocity space is discretized using appropriate quadratures (see Section~\ref{sec:quadratures}), resulting in a DVM approach where the problem is converted to $N$ equations with fixed velocity $\mathbf v \in (\mathbf v_i)_{i \in [1,N]}$. Then, one integrates the collision operator of the BGK equation~\eqref{eq:CollOp} with a Crank-Nicolson scheme and applies the midpoint rule to the flux $\mathcal F=\mathbf v \cdot \frac{\partial f}{\partial \mathbf x}$:
\begin{equation}
 f(\mathbf x,\mathbf v, t + \Delta t) = f(\mathbf x,\mathbf v, t)  - \Delta t \mathcal F(\mathbf x, \mathbf v, t+\Delta t/2) + \frac{\Delta t}{2}\left[\Omega(\mathbf x,\mathbf v, t + \Delta t)+\Omega(\mathbf x,\mathbf v, t)\right].
\end{equation}
By cleverly rearranging the terms and introducing the two new variables
\begin{eqnarray}
\tilde{f} &= f - \frac{\Delta t}{2}\Omega = \frac{2\tau+\Delta t}{2\tau}f-\frac{\Delta t}{2\tau}f^t \label{eq:twoCN1} \\
\textrm{and} \quad \hat{f} &= f + \frac{\Delta t}{2}\Omega = \frac{2\tau-\Delta t}{2\tau}f+\frac{\Delta t}{2\tau}f^t,
\label{eq:twoCN2}
\end{eqnarray}
with $\tau=1/\nu$ being the relaxation time, it is possible to eliminate the stiffness of the collision operator by treating it implicitly:
\begin{equation}
\tilde{f}(\mathbf x, \mathbf v, t+\Delta t)=\hat{f}(\mathbf x, \mathbf v, t) - \Delta t \mathcal F(\mathbf x, \mathbf v,t+\Delta t/2).
\end{equation}
With a suitable construction of the flux term $\mathcal F$, it is thus possible to achieve an implicit integration of the coupled advection and relaxation. However, a time step of this scheme is completed by using the relation
\begin{equation}
  \hat{f} = \frac{2\tau - \Delta t}{2\tau + \Delta t} \tilde{f} + \frac{2\Delta t}{2\tau + \Delta t} f^t,
\end{equation}
where the pre-factor $\frac{2\tau - \Delta t}{2\tau + \Delta t}$ multiplying the distribution function $\tilde{f}$ becomes a problem for stochastic particle methods. Indeed, since it can become negative at large time steps relative to the relaxation time and because this pre-factor would be interpreted as a probability for individual particles (see~\citet{pfeiffer2022exponential}), this leads to major problems in the adaptation to particle methods.

In the following, the construction of an integration method with unconditionally positive pre-factors, which are also always limited to $[0;1]$, is described. It can then be used directly as a DVM, but also as a stochastic particle method.

\subsection{Exponential time differencing BGK}

In order to construct a second-order multiscale DVM, the stiffness induced in the BGK equation when the relaxation frequency $\nu$ increases needs to be addressed. The approach followed here is the same as used in~\citet{pfeiffer2022exponential}, where particular attention is paid to the positivity of pre-factors in order to build a particle-based solver. An exact integration of the BGK equation is first conducted using exponential time differencing, which is particularly advantageous to handle a stiff linear term in ordinary differential equations~\cite{cox2002exponential}:
\begin{equation}
 f(t^{n+1}) = f(t^n)e^{-\nu \Delta t}  + e^{-\nu \Delta t}\int_0^{\Delta t}e^{\nu s} \left( \nu f^t(t^n+s) - \mathbf v \frac{\partial f}{\partial \mathbf x}(t^n+s)\right)\,ds.
\label{eq:exactEI}
\end{equation}
To deal with the remaining nonlinear stiff term $\nu f^t$ in the integral, a simple linear approximation is used:
\begin{equation}
f^t(t+s) = f^t(t)+ \frac{s}{\Delta t}(f^t(t+\Delta t)-f^t(t)) + O(s^2) + O(s\Delta t).
\label{eq:linearapprox}
\end{equation}
This leads to a Crank-Nicolson-type scheme and preserves the second-order accuracy.

To build a finite volume solver, as commonly chosen for DVM-based methods in order to benefit from the conservation properties and properly handle discontinuities, the equation is then integrated on a control volume $V_j$ centered on point $\mathbf x_j$. Therefore, in the following, we use the notations
\begin{eqnarray}
f_{j,n}&=&\frac{1}{|V_j|}\int_{V_j} f(\mathbf x, \mathbf v, t^n) d\mathbf x \quad \text{and} \\
F_{j,n+1/2} &=& \frac{1}{|V_j|}\int_{V_j} \mathbf v \cdot \frac{\partial f}{\partial \mathbf x}(\mathbf x, \mathbf v, t^n+\frac{\Delta t}{2})d\mathbf x.
    \label{eq:fvnotations}
\end{eqnarray}

The space dependency is omitted in the following where it is not explicitly required. With~\eqref{eq:linearapprox} and using a midpoint rule for the time integration of the flux, Eq.~\eqref{eq:exactEI} becomes
\begin{eqnarray}
f_{n+1} &=& f_n e^{-\nu \Delta t} - \gamma \Delta t F_{n+1/2} \nonumber \\
&& + e^{-\nu \Delta t}\left[ f^t_n\left( \frac{e^{\nu \Delta t}}{\nu \Delta t} - 1 -\frac{1}{\nu \Delta t}\right)+f^t_{n+1}\left( e^{\nu \Delta t} + \frac{1}{\nu \Delta t} - \frac{e^{\nu \Delta t}}{\nu \Delta t}\right)\right] \nonumber \\
&=& f_ne^{-\nu \Delta t} - \gamma \Delta t F_{n+1/2}+ \left(1-e^{-\nu \Delta t}\right)\times \left(Af^t_n + Bf^t_{n+1}\right),
\label{eq:linexp}
\end{eqnarray}
with
\begin{equation}
\gamma=\frac{1-e^{-\nu\Delta t}}{\nu\Delta t},\quad
A = \frac{1}{\nu \Delta t} - \frac{e^{-\nu\Delta t}}{1-e^{-\nu\Delta t}}\quad\textrm{and} \quad B = \frac{1}{1-e^{-\nu\Delta t}}-\frac{1}{\nu\Delta t}.
\end{equation}

Next, in a similar way as in DUGKS~\cite{guo2013discrete}, two modified distributions can be introduced :
\begin{eqnarray}
\hat{f} &=&  \frac{1}{\gamma} \left( f e^{-\nu \Delta t} + (1-e^{-\nu \Delta t})Af^t \right)
\label{eq:fhat}
\\ \textrm{and} \quad
\widetilde{f} &=&  \frac{1}{\gamma} \left( f - (1-e^{-\nu \Delta t})Bf^t \right).
\label{eq:ftilde}
\end{eqnarray}
Using these notations and dividing~\eqref{eq:linexp} by $\gamma$ then results in a simple scheme where the advection and relaxation processes actually remain coupled in a second-order accurate time integration:
\begin{equation}
    \widetilde{f}_{n+1} = \hat{f}_n - \Delta t F_{n+1/2}.
    \label{eq:update}
\end{equation}
Thus, the implicit part of the BGK operator can be handled by tracking the new distribution function $\widetilde{f}$. And by combining~\eqref{eq:fhat} and~\eqref{eq:ftilde}, $\hat{f}$ can be constructed from $\widetilde{f}$ for the next time step:
\begin{equation}
\hat{f} = e^{-\nu \Delta t} \widetilde{f} + (1-e^{-\nu \Delta t})f^t.
\label{eq:finallincom}
\end{equation}
In this equation, both pre-factors are always positive, in the interval $[0;1]$ and sum up to 1.
The target distribution $f^t$ can be obtained at every time step using the macroscopic moments of $f$, which directly correspond to those of $\widetilde{f}$ (see Section~\ref{sec:moments}). The original distribution function can then easily be obtained by
\begin{eqnarray}
    f&=&\gamma \widetilde{f} + (1-e^{-\nu\Delta t})Bf^t \nonumber \\
    &=& \gamma\widetilde{f} + (1- \gamma)f^t.
\label{eq:ffromftilde}
\end{eqnarray}
Note that, here as well, both pre-factors are always in $[0;1]$ and sum up to 1 because $\nu\Delta t$ is obviously positive.

The positivity of all factors in equations \eqref{eq:fhat}, \eqref{eq:finallincom} and \eqref{eq:ffromftilde} prevent any negative values to appear in the solution $f$ during these steps, as long as the target distribution $f^t$ is also positive. The conservation of positivity during the update described by equation \eqref{eq:update} depends on the construction of a stable flux term. This is discussed with more details in~\ref{apx:positive}.

\subsection{Flux evaluation}
\label{sec:flux}

The remaining critical part is the evaluation of the flux $F_{n+1/2}$, for which the value of the distribution on the interfaces between mesh cells at time $t_n+\Delta t/2$ is needed. The BGK equation is therefore integrated again with exponential differencing, but this time along a characteristic line that ends at a point $\mathbf x_k$ on the cell interface to finally reconstruct $f_{n+1/2}(\mathbf x_k)$ and thus calculate the flux. Along this line ($\mathbf x_k + \mathbf v t$), the BGK equation becomes simply
\begin{equation}
  \frac{d f}{d t} = \nu (f^t-f).
\end{equation}
Using modified distribution functions again to integrate within a half time step, this results in
\begin{equation}
    \widetilde{f}^r\left(\mathbf x_k, \mathbf v, t_n+\frac{\Delta t}{2}\right) = \hat{f}^r\left(\mathbf x_k - \mathbf v \frac{\Delta t}{2}, \mathbf v, t_n\right)
    \label{eq:halfstep}
\end{equation}
with
\begin{eqnarray}
\hat{f}^r &=&  \frac{1}{\gamma^r} \left( f e^{-\nu \frac{\Delta t}{2}} + (1-e^{-\nu \frac{\Delta t}{2}})A^rf^t \right)
\label{eq:fhat^r}
\\ \textrm{and} \quad
\widetilde{f}^r &=&  \frac{1}{\gamma^r} \left( f - (1-e^{-\nu \frac{\Delta t}{2}})B^rf^t \right)
\label{eq:ftilde^r}
\end{eqnarray}
where
\begin{equation}
\gamma^r=\frac{1-e^{-\nu\frac{\Delta t}{2}}}{\nu\frac{\Delta t}{2}}\quad
A^r = \frac{1}{\nu \frac{\Delta t}{2}} - \frac{e^{-\nu\frac{\Delta t}{2}}}{1-e^{-\nu\frac{\Delta t}{2}}}\quad
\textrm{and} \quad B^r = \frac{1}{1-e^{-\nu\frac{\Delta t}{2}}}-\frac{1}{\nu\frac{\Delta t}{2}}.
\end{equation}

 $\hat{f}^r$ is obtained directly from $\widetilde{f}$ by combining equations~\eqref{eq:fhat^r} and~\eqref{eq:ffromftilde}:
 \begin{equation}
  \hat{f}^r= \frac{\gamma}{\gamma^r}e^{-\nu\frac{\Delta t}{2}}\widetilde{f} + \left(1-\frac{\gamma}{\gamma^r}e^{-\nu\frac{\Delta t}{2}}\right)f^t.
  \label{eq:ftildetofhat^r}
 \end{equation}
 $\hat{f}^r$ then needs to be reconstructed from its value at the cell center to the point $\mathbf x_k - \mathbf v \frac{\Delta t}{2}$ with a linear approximation, this time to preserve the second-order spatial accuracy:
\begin{equation}
    \hat{f}^r\left(\mathbf x_k - \mathbf v \frac{\Delta t}{2},\mathbf v, t_n\right) = \hat{f}^r(\mathbf x_j, \mathbf v, t_n) + \left(\mathbf x_k - \mathbf v \frac{\Delta t}{2} - \mathbf x_j\right) \nabla \hat{f}^r(\mathbf x_j, \mathbf v, t_n),
    \label{eq:reconstruction}
\end{equation}
where $\mathbf x_j$ is the center of the cell adjacent to the the interface, chosen in the usual upwind manner, i.e. $\mathbf v\cdot (\mathbf x_k - \mathbf x_j) > 0$.

As this scheme is meant to be used on unstructured meshes, the gradients $\nabla \hat{f}^r$ are calculated from cell-averaged values in neighbouring cells with the least-squares approach used in \citet{zhu2016discrete}.
To prevent numerical instabilities from appearing in shock regions, a gradient limiter can be applied, following the method of~\citet{barth1989design}. For even better convergence when simulating flows with strong discontinuities, the~\citet{venkatakrishnan1995convergence} limiter was also implemented.

Using a Stokes formula, the numerical flux is then computed for each cell $V_j$ by summing up contributions from the interfaces $S_k$, with center of mass $\mathbf x_k$ and outward-pointing normal vector $\mathbf n_k$:
\begin{eqnarray}
  F_{j,n+1/2}&=&\sum_{S_k  \in \partial V_j} \frac{|S_k|}{|V_j|}(\mathbf n_k \cdot \mathbf v) f(\mathbf x_k, \mathbf v, t_n+\frac{\Delta t}{2}) \nonumber \\
  &=&\sum_{S_k  \in \partial V_j} \frac{|S_k|}{|V_j|}(\mathbf n_k \cdot \mathbf v) \left(\gamma^r\widetilde{f}^r(\mathbf x_k, \mathbf v, t_n+\frac{\Delta t}{2})+(1-\gamma^r)f^t_{n+1/2}(\mathbf v)\right),
  \label{eq:sumflux}
\end{eqnarray}
where $f^t_{n+1/2}$ is constructed from the moments of $\widetilde{f}^r\left(\mathbf x_k, \makebox[1ex]{\textbf{$\cdot$}}, t_n+\frac{\Delta t}{2}\right)$.

\subsection{Properties of the scheme}

The resulting exponential differencing DVM (ED-DVM) does not rely on a splitting of the advection and relaxation processes, meaning that the mean free path of the particles does not need to be resolved, allowing for much larger mesh cells than those required with Discrete Simulation Monte Carlo (DSMC)  in near-continuum cases. The second-order convergence in space and time, proved analytically in~\ref{apx:order}, also improves accuracy compared to the particle-based methods, usually only reaching a maximum convergence order of 1. Besides, the implicit integration of the relaxation term allows for the time step to be chosen larger than the relaxation time. However, due to the overall explicit nature of the method, the Courant-Friedrichs-Lewy (CFL) condition still needs to be fulfilled.

The devised ED-DVM is in fact an asymptotic preserving method, therefore able to solve the BGK equation in all regimes with second-order accuracy. Indeed, in the free molecular flow limit ($\frac{1}{\nu}=\tau\gg \Delta t$), all rescaled distribution functions are equivalent to the original $f$ and the scheme then solves a simple advection equation for each velocity point, retaining second-order finite volume discretization. Besides, in the continuum limit ($\tau\ll\Delta t$), it can be shown that a time-dependent Chapman-Enskog distribution function is retrieved at the cell boundaries for the flux calculation, turning the ED-DVM into a consistent second-order Navier-Stokes solver.
The proof of the asymptotic behavior of the method can be found with more details in~\ref{apx:AP}.

\begin{figure}
  \centering
  \includegraphics{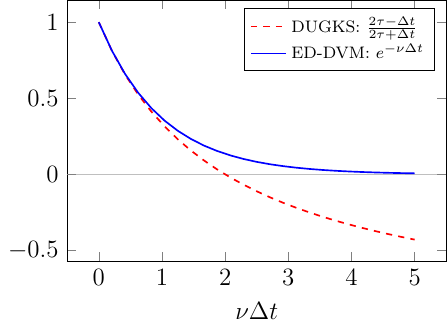} 
  \caption{Pre-factor in front of $\widetilde{f}$ in the equation to build $\hat{f}$.}
  \label{fig:prefactors}
\end{figure}

Furthermore, the conservative property of the BGK operator and the use of a finite volume scheme ensure that density, momentum, and energy are conserved up to the numerical error introduced by the velocity discretization. The discretized collision operator is not conservative but properly choosing the velocity quadrature reduces the induced numerical deviation, which has therefore not been a problem in the following simulations. A fully conservative scheme could be implemented following the idea of~\citet{mieussens2000discrete}, but this would come with the additional cost of solving a system of at least 5 nonlinear equations in every mesh cell at every time step.

The main difference between ED-DVM and DUGKS is the aforementioned positivity of the pre-factors used in the "rescaling" that combines linearly the distribution function $f$ with the target $f^t$. This property, in addition to ensuring the positivity of the distribution function as long as the target distribution and flux term are also positive, is what allows the direct extension of this DVM to a stochastic particle scheme. The factors regarded as probabilities in such a method~\cite{pfeiffer2022exponential} would mainly be those in Eq.~\eqref{eq:finallincom} and they indeed remain between 0 and 1 independently of the relaxation frequency and time step, which is not the case with their DUGKS equivalents, as displayed on Fig.~\ref{fig:prefactors}.

\section{Numerical implementation}
\label{sec:implementation}

\subsection{Discrete velocity quadratures}
\label{sec:quadratures}

Focusing now on the velocity discretization, the choice of the quadrature points and weights depends on the problem considered. The commonly used discretizations are based either on the Gauss-Hermite quadrature or on composite Newton-Cotes rules~\cite{yang1995rarefied,wang2018comparative,guo2013discrete}. The quadrature rules are given here in one dimension but can be simply extended to two or three dimensions by discretizing the additional velocity directions in the same manner and multiplying the respective weights.

For flows that can be considered almost isothermal (around a reference temperature $T_{GH}$) and with no region where the particle distribution is too far from thermal equilibrium, a Gauss-Hermite quadrature can be used. With $h_{N,i}$ the $N$ roots of the Hermite polynomial $H_N$, the velocity discretization points and weights are
\begin{equation}
  v_i = \sqrt{2RT_{GH}}h_{N,i},\quad w_i= \sqrt{2RT_{GH}}\frac{2^{N-1}N!\sqrt{\pi}}{N^2[H_{N-1}(h_{N,i})]^2}e^{{h_{N,i}}^2}.
\end{equation}
This choice of quadrature allows for an accurate representation of distribution functions with a limited number of points, provided that the deviation from equilibrium is small, and ensures that the BGK relaxation operator is conservative in the purely isothermal case \cite{guo2021progress}.
 The precision of this quadrature can be improved by increasing the polynomial degree $N$ and therefore the number of velocity points.

In cases where strong non-equilibrium effects should be observed (e.g. shocks, hypersonic or highly rarefied flows), a uniform choice of discretization, without assuming a Gaussian shape for the distribution, is preferable. A Newton-Cotes quadrature is therefore used to discretize the velocity space for such simulations, spanning a range $[v_{min},v_{max}]$ with evenly spaced points. Concerning the quadrature weights, since high order Newton-Cotes formulas do not necessarily increase the interpolation accuracy, we divide $[v_{min},v_{max}]$ into smaller equal intervals in which low order formulas are used, while ensuring that this results in an $N$ point composite quadrature.

\subsection{Calculation of the moments}
\label{sec:moments}

In order to reduce the computational cost when simulating a $d$-dimensional problem with $d=1$ or $2$, reduced distribution functions are introduced, as commonly done~\cite{chu1965kinetictheoretic,yang1995rarefied} to avoid discretizing the velocity space in the unnecessary directions:
\begin{eqnarray}
  g(\mathbf x, \mathbf v',t) &=& \int f(\mathbf x, \mathbf v', \bm{\eta},t) d\bm{\eta} \\ \text{and} \quad h(\mathbf x, \mathbf v',t) &=& \int \bm{\eta}^2 f(\mathbf x, \mathbf v', \bm{\eta},t) d\bm{\eta}
\end{eqnarray}
where, setting the $x, y$ and $z$ velocity components as $\mathbf v=(\xi_1,\xi_2,\xi_3)^T$, we write $\mathbf v'=(\xi_1,...,\xi_d)^T$ and $\bm{\eta}=(\xi_{d+1},...,\xi_3)^T$.
For the Shakhov model, the corresponding target distributions are then:
\begin{eqnarray}
  g^S(\mathbf v')&=&\int f^S(\mathbf v) d\bm{\eta}\nonumber\\
  &=&g^M\left[1+(1-\textrm{Pr})\frac{\mathbf c' \cdot \mathbf q}{5\rho(RT)^2}\left(\frac{\mathbf c'^2}{RT}-d-2\right)\right] \\
  h^S(\mathbf v')&=&\int \bm{\eta}^2 f^S(\mathbf v) d\bm{\eta}\nonumber\\
  &=& g^M(3-d)RT\left[1+(1-\textrm{Pr})\frac{\mathbf c' \cdot \mathbf q}{5\rho(RT)^2}\left(\frac{\mathbf c'^2}{RT}-d\right)\right]
\end{eqnarray}
with $g^M=\frac{\rho}{(2\pi RT)^{d/2}}e^{-\frac{\mathbf c'^2}{2RT}}$ and $\mathbf c'=\mathbf v' - \mathbf u$.

The evolution of these reduced distributions is also obtained from the BGK equation by simply replacing $f$ with $g$ or $h$, and $f^S$ with $g^S$ or $h^S$. Therefore, the whole scheme detailed above is applied simultaneously to $g$ and $h$, and the moments used to construct the target distributions $g^S$ and $h^S$ are directly computed through discrete sums:
\begin{equation}
  \mathbf W = \begin{pmatrix} \rho \\ \rho \mathbf u \\ \rho E \end{pmatrix} = \sum_{i=1}^N w_i \left[\bm{\psi} (\mathbf v'_i) g(\mathbf x,\mathbf v'_i, t)+\bm{\zeta}h(\mathbf x,\mathbf v'_i, t)\right]
  \label{eq:moments}
\end{equation}
with $\bm{\zeta}=(0,0,1)^T$ and $\bm{\psi}=(1,\mathbf v,\mathbf v^2/2)^T.$

Besides, as the first moments $\mathbf W$ of a distribution $f$ and its target counterpart $f^t$ are the same, those of $\widetilde f$ are also equal to $\mathbf W$, because the factors of the linear combination in~\eqref{eq:ftilde} sum up to 1. Eq.~\eqref{eq:moments} can therefore easily be rewritten using the distributions $\widetilde{g}$ and $\widetilde{h}$, rescaled in this same way, which are the functions actually available during a time step of the method.

However, higher moments are not conserved by target distributions. The heat flux vector of the Shakhov distribution is indeed $\mathbf q(f^S) = (1-\mathrm{Pr})\mathbf q(f)$, altering $\widetilde{\mathbf q}= \mathbf q(\widetilde f)$ obtained through
\begin{equation}
  \label{eq:heatflux1}
  \widetilde{\mathbf{q}}=\frac{1}{2}\sum_{i=1}^N w_i \mathbf c'_i\left[\mathbf c'^2_i\widetilde{g}(\mathbf x,\mathbf v'_i, t)+\widetilde{h}(\mathbf x,\mathbf v'_i, t)\right].
\end{equation}
Since the Shakhov model requires the physical heat flux $\mathbf q = \mathbf q(f)$, the virtual one obtained from $\widetilde f$ needs to be rescaled using Eq.~\eqref{eq:ffromftilde}:
\begin{equation}
  \label{eq:heatflux2}
  \mathbf q = \gamma\widetilde{\mathbf q} + (1-\gamma)(1-\mathrm{Pr}) \mathbf q = \frac{\gamma}{\mathrm{Pr}+\gamma(1-\mathrm{Pr})}\widetilde{\mathbf q}
\end{equation}
The pressure tensor is also retrieved from $\widetilde{\mathcal{P}}=\mathcal P(\widetilde f)$ via
\begin{equation}
  \label{ptShakhov}
  \mathcal P = \gamma \widetilde{\mathcal P} + (1-\gamma)\rho RT \mathcal{I}.
\end{equation}

The same reasoning can be applied if the ESBGK model is used, yielding:
\begin{eqnarray}
  g^{ES}(\mathbf v')&=&\frac{\rho}{\sqrt{\det_d(2\pi\mathcal A')}}\exp{\left[-\frac{\mathbf c'^T \mathcal A'^{-1} \mathbf c'}{2}\right]} \\
  h^{ES}(\mathbf v')&=&\frac{\rho}{\sqrt{\det_d(2\pi\mathcal A')}}\exp{\left[-\frac{\mathbf c'^T \mathcal A'^{-1} \mathbf c'}{2}\right]} \Tr(\mathcal A'')\\
  \text{where}&& \mathcal A'_{ij}=\mathcal A_{ij}\ \text{for}\ i,j\leq d\ \text{and}\ \mathcal A''_{ij}=\mathcal A_{ij}\ \text{for}\ i,j> d\nonumber\\
  \mathcal P &=&\frac{1}{\frac{1}{\mathrm{Pr}}+\gamma(1-\frac{1}{\mathrm{Pr}})}\left(\gamma \widetilde{\mathcal P}+(1-\gamma)\frac{\rho RT}{\mathrm{Pr}}\mathcal{I}\right)\\
  \mathbf q &=& \gamma\widetilde{\mathbf q}.
\end{eqnarray}

\subsection{Boundary conditions}
\label{sec:boundaries}

Two types of boundary conditions have been implemented in this study for gas/wall interactions: specular and diffusive reflection. In both cases, the first step is the construction of a boundary-specific distribution function on the wall. In order to validate these boundary conditions, a single one is applied at each wall in the following test cases, but for more realistic wall interactions, a linear combination of the specular and diffusive wall distributions could be used.

The specular reflection distribution is the same as the inner cell distribution, but obtained after perfectly reflecting the velocities directed onto the wall:
\begin{equation}
    \forall \mathbf v \text{ so that } \mathbf v \cdot \mathbf n_w < 0,\quad f(\mathbf x_w, \mathbf v, t) = f(\mathbf x_w, \mathbf v - 2(\mathbf v \cdot\mathbf n_w)\mathbf n_w, t).
\end{equation}
This boundary condition is only applicable to a static wall with a normal vector $\mathbf{n}_w$ aligned with the velocity grid, so that for a grid point $\mathbf{v}_i$, the velocity $\mathbf v_i - 2(\mathbf v_i \cdot\mathbf n_w)\mathbf n_w$ is another grid point. If more complex geometries have to be simulated with this type of boundaries, a velocity interpolation could be used~\cite{dechriste2014methodes}.

The diffusive reflection distribution is a simple Maxwellian $f^M_w$ with wall temperature and velocity, first calculated with arbitrary density $\rho_\text{tmp}$.
The actual wall density is then adjusted to conserve the number of particles hitting the wall~\cite{cercignani1988boltzmann}:
\begin{equation}
  \rho_w = - \rho_\text{tmp} \times \left[\sum_{\mathbf v\cdot \mathbf n_w > 0} (\mathbf v\cdot \mathbf n_w) f^M_w\right]^{-1} \times \sum_{\mathbf v\cdot \mathbf n_w < 0} (\mathbf v\cdot \mathbf n_w) f(\mathbf x_w, \mathbf v, t).
\end{equation}

In order to preserve the second-order accuracy at the boundaries, ghost cells are created outside of the domain to be able to compute gradients and perform the reconstruction step in the boundary cells. In these ghost cells, the value of the distribution function is interpolated from inner and wall values while ensuring compatibility with the reconstruction algorithm, following the method described in~\citet{baranger2019numerical}. The finite volume flux can then be calculated as inside of the domain.

\subsection{Algorithm}
\label{sec:algo}

The resulting ED-DVM algorithm consists of updating the distribution function $\widetilde{f}_n$ in each of the $N_{\mathrm{c}}$ mesh cell at each time step. The time step is chosen in accordance with the CFL condition, given by:
\begin{equation}
  C(\Delta t) = \Delta t \max_{j=1,N_{\mathrm{c}}}\left[\max_{i=1,N}\left(\sum_{S_k  \in \partial V_j} \frac{|S_k|}{|V_j|}\max(0,\mathbf n_k \cdot \mathbf v_i)\right)\right]\leq 1.
\end{equation}
This numerical scheme is implemented in the open-source PIC-DSMC code PICLas~\cite{fasoulas2019combining} and can be summarized as follows:
\begin{enumerate}
  \item Compute the target distribution $f^t_n$ using the macroscopic values obtained via the moments of $\widetilde{f}_n$ (Eq.~\ref{eq:moments} to~\ref{eq:heatflux2}).
  \item Compute the distribution function $\hat{f}^r_n$ from $\widetilde{f}_{n}$ (Eq.~\ref{eq:ftildetofhat^r}).
  \item Compute the necessary distributions at the boundaries, depending on the boundary condition (Section \ref{sec:boundaries}).
  \item Compute the spatial gradients of $\hat{f}^r_n$ and apply a limiter if necessary (Section \ref{sec:flux}).
  \item Reconstruct $\widetilde{f}^r_{n+1/2}$ at the cell interfaces (Eq.~\ref{eq:reconstruction} and~\ref{eq:halfstep}).
  \item Sum up the flux contributions at interfaces and boundaries (Eq.~\ref{eq:sumflux}).
  \item Compute the distribution function $\hat{f}_n$ from $\widetilde{f}_{n}$ (Eq.~\ref{eq:finallincom}).
  \item Update the distribution function to $\widetilde{f}_{n+1}$ (Eq.~\ref{eq:update}).
\end{enumerate}

For the first time step of a simulation, initialized with distribution $f_0$, there is no need to determine $\widetilde{f}_0$, even if $f_0$ is a non-equilibrium state. Indeed, moments can be directly computed from the initial state without any rescaling. Similarly, $\hat f^r_0$ in step 2 and $\hat f_0$ in step 7 should be directly computed from $f_0$ using equation~\eqref{eq:fhat^r} instead of~\eqref{eq:ftildetofhat^r} and equation~\eqref{eq:fhat} instead of~\eqref{eq:finallincom}.

The DUGKS method is also implemented in the same code, using the same framework (velocity quadratures, boundaries, flux reconstruction). Both DVMs make up a new module of PICLas, which can be compiled instead of the particle solver that already comprises the DSMC, BGK and Fokker-Planck particle methods. This new feature allows for comparisons between stochastic and deterministic methods, performance gains in regimes where the particle solver leads to strong statistical noise, and the possibility to implement a coupling of both approaches.

\section{Results}
\label{sec:results}

\subsection{Heat flux relaxation}

\begin{figure}
  \centering
  \subfloat[ED-DVM]{\includegraphics{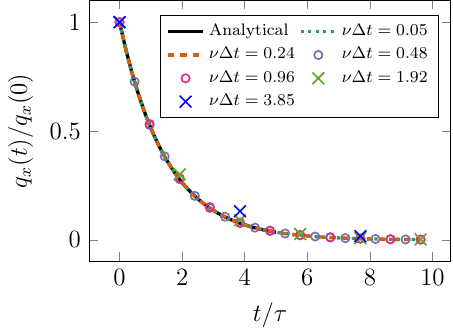}} 
  \hfill
  \subfloat[DUGKS]{\includegraphics{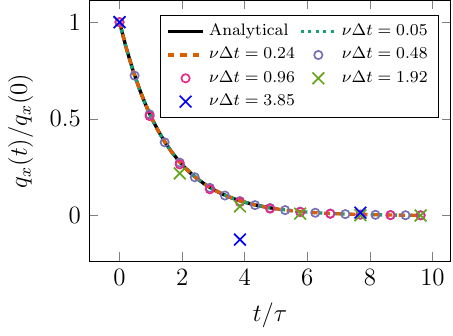}} 
  \caption{Heat flux relaxation with different time step sizes $\Delta t$.}
  \label{fig:heatflux}
\end{figure}

\begin{figure}
  \centering
  \includegraphics{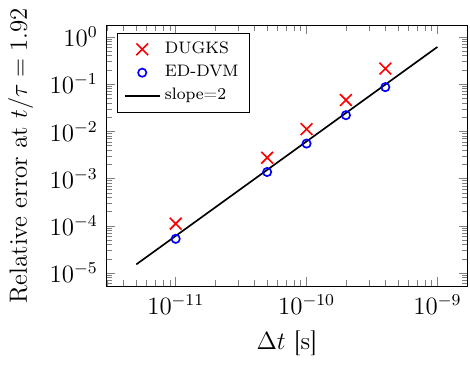}
  \caption{Heat flux relaxation: error convergence for ED-DVM and DUGKS.}
  \label{fig:heatflux_error}
\end{figure}

As a first test case, a spatially homogeneous heat flux relaxation is considered, in order to validate the second-order accuracy of the ED-DVM scheme in time. A single adiabatic mesh cell is filled with argon gas at density $\rho=1.79$ kg/m$^3$ and temperature $T=273$~K. Only one velocity direction is allowed and the heat flux is initially set in that direction to $q_x(0)=\frac{1}{2}\rho(RT)^{3/2}$ using Grad's 13 moments distribution~\cite{struchtrup2003regularization} with a traceless pressure tensor equal to zero:
\begin{equation}
  f_{init}=f^M\left[1-\frac{\mathbf q \cdot\mathbf c}{\rho R^2 T^2}\left(1-\frac{\mathbf c^2}{5 RT}\right)\right].
\end{equation}
Using the Shakhov BGK model, the heat flux then relaxes exponentially towards equilibrium at a rate $\mathrm{Pr}/\tau$. Fig.~\ref{fig:heatflux} shows that ED-DVM is able to properly resolve this test case, even with time steps several times larger than the relaxation time for which DUGKS appears more unstable. Furthermore, the error comparison (Fig.~\ref{fig:heatflux_error}) clearly displays the expected second-order convergence in time, with ED-DVM performing slightly better than DUGKS.

\subsection{Double Maxwellian relaxation}

\begin{figure}
  \subfloat[Initialization]{\includegraphics{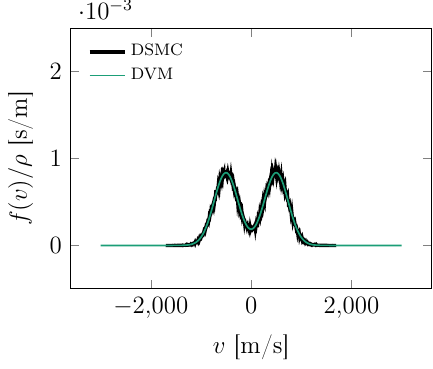}} 
  \hfill
  \subfloat[After $\Delta t = 28\tau=10^{-4}$ s]{\includegraphics{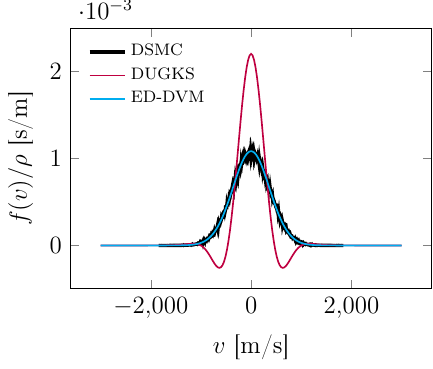}} 
  \caption{Relaxation of a sum of two Maxwellian distributions.}
  \label{fig:doubledist}
\end{figure}

To better illustrate the difference between ED-DVM and DUGKS, a test case initialized in a strong non-equilibrium state is considered. As with the previous test case, a single adiabatic mesh cell is filled with argon gas, this time with a velocity distribution corresponding to the sum of two 1D Maxwellian distributions, each at the same temperature of 273 K but different mean velocities: $u_{x,1}=-500$ m/s and $u_{x,2}=500$ m/s. This initial velocity distribution, displayed on Figure \ref{fig:doubledist}a, is then left to relax for $10^{-4}$ s in a single time step for both DVMs (DUGKS and ED-DVM), while the same is done with DSMC over 1000 time steps to resolve the collision frequency.
The ESBGK model is used to ensure that negative values of the solution are only due to the solver and not the BGK model itself, because, in contrast to the Shakhov model, the ESBGK target distribution is unconditionally positive. 201 evenly distributed velocity points between -3000 and 3000 m/s were used to obtain a visually smooth solution. The density is set so as to give rise to a high relaxation factor ($\nu\Delta t \simeq 28$).

The results of this test case are displayed on Figure \ref{fig:doubledist}b, showing that ED-DVM relaxes towards the correct equilibrium distribution, corresponding to a Maxwellian centered around 0 m/s, also reached by the DSMC simulation. On the other hand, due to the aforementioned negative factors in its formulation, DUGKS fails to reproduce this equilibrium distribution in this timeframe. Even in such a case without flux terms, the local operations on the distribution function result in a non-positive solution for DUGKS. The conservative moments are still correct in all DVM simulations as the negative distribution values can be considered in the DVM framework, allowing for simulations that can remain stable even when such cases occur, but it is clear that such a solution has lost its physical meaning when it comes to the statistical description of the gas.

\begin{figure}
  \subfloat[DUGKS]{\includegraphics{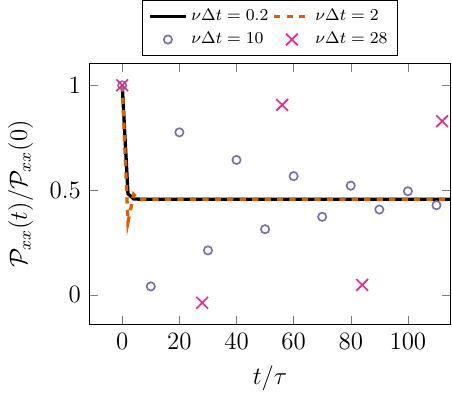}} 
  \hfill
  \subfloat[ED-DVM]{\includegraphics{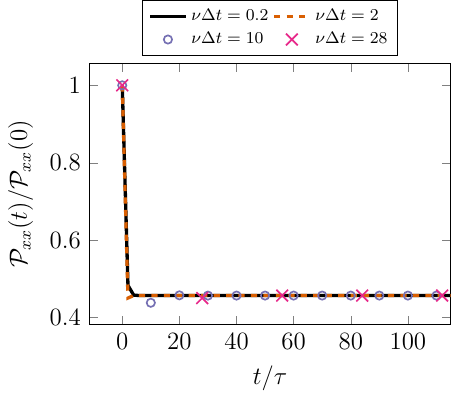}} 
  \caption{Relaxation of the pressure tensor of a sum of two Maxwellian distributions.}
  \label{fig:dd_relax}
\end{figure}

For a better understanding of this stability issue, the simulation is conducted for a longer time and the relaxation of a diagonal entry of the pressure tensor (the first non-conservative moment) is displayed for both methods on Figure~\ref{fig:dd_relax}. A comparison is made there using smaller time steps, and it is thus clear that ED-DVM is better than DUGKS at capturing the right relaxation time in this case. Although DUGKS should eventually reach equilibrium, the oscillating convergence of the solution on time scales larger than the relaxation time, already clear with $\nu\Delta t = 10$, is what prevents DUGKS from providing a physically meaningful time-accurate solution.

However, in addition to the high relaxation factor, a distribution function very different from the local equilibrium is also required to obtain such a problematic situation. It has therefore not appeared in the validation cases presented thereafter, unless provoked on purpose (see Section~\ref{sec:cylinder}). Indeed, even in cases where $\nu\Delta t > 2$, where DUGKS factors are already negative, the limited difference between $f$ and $f^t$ often ensures the positivity of all "rescaled" distributions.

\subsection{Sod shock tube}

The first 1D test case is the widely studied Sod shock tube problem, where a tube of length $L=1$ m contains a fluid, here argon, with an initial discontinuity in density and temperature at $x=0.5$ m. The left and right temperatures are set to $T_l=273$K and $T_r=218.4$K while the initial density is chosen depending on the flow's Knudsen number, with the condition $\rho_r=0.125\rho_l$. The Knudsen number Kn is defined as the ratio of the mean free path $\lambda$ to the characteristic length $L$ of the problem, ie.

\begin{equation}
  \text{Kn} = \frac{\lambda}{L} = \frac{\mu}{\rho L} \sqrt{\frac{\pi}{2RT}}.
\end{equation}

The tube was discretized in 100 equal cells and two different regimes were investigated: a dense case with $\rho_r=1.0725\times 10^{-5}$ kg/m$^3$, resulting in a maximum Knudsen number around Kn = $0.01$ and a rarefied regime with $\rho_r=1.0725\times 10^{-7}$ kg/m$^3$ for a maximum of Kn$=1$. To solve both cases with ED-DVM and in order to respect the CFL condition with a one-dimensional CFL number of $\max(|v_{min}|,|v_{max}|)\frac{\Delta t}{ \Delta x}=0.9$, the time step is set to $\Delta t = 6.667\times 10^{-6}$ s and the uniform velocity space limited to $[-4\sqrt{2RT_l};4\sqrt{2RT_l}]$. The~\citet{barth1989design} limiter was used in order to avoid any nonphysical oscillations due to the high gradients around $x=0.5$ m. The shock then propagates for $7\times 10^{-4}$ s and results for the first three moments are compared at that point in time to those obtained via DSMC.

\begin{figure*}
\begin{adjustwidth}{-0.8cm}{-0.8cm}
  \centering
  \subfloat{\scalebox{0.7}{\includegraphics{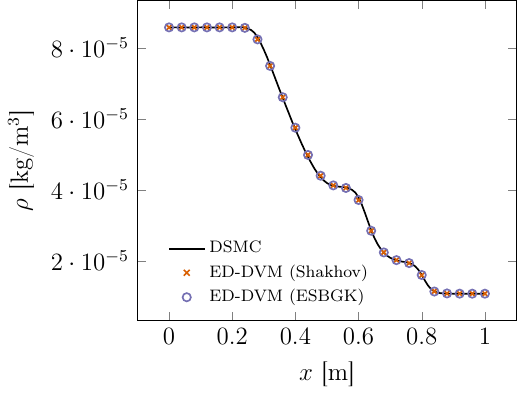}}} 
  \hfill
  \subfloat{\scalebox{0.7}{\includegraphics{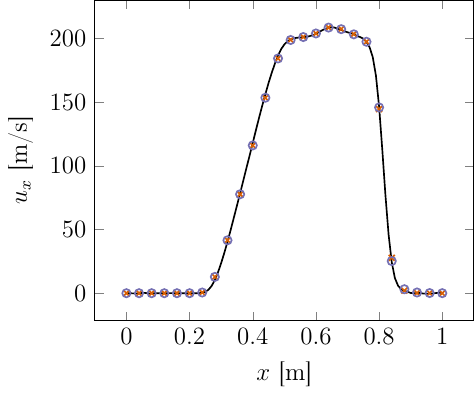}}} 
  \hfill
  \subfloat{\scalebox{0.7}{\includegraphics{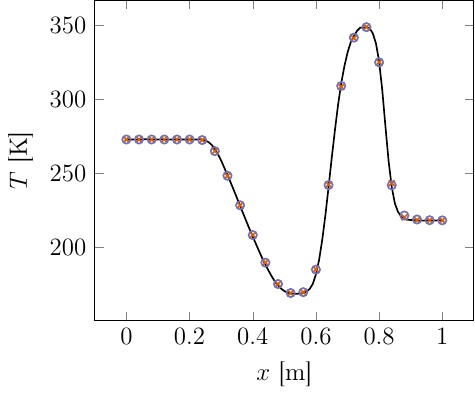}}} 
\end{adjustwidth}
  \caption{Sod shock tube at Kn $=0.01$: density, velocity and temperature profiles at $t=7\times 10^{-4}$ s.}
  \label{fig:soddense}
\end{figure*}

\begin{figure*}
\begin{adjustwidth}{-0.8cm}{-0.8cm}
  \centering
  \subfloat{\scalebox{0.7}{\includegraphics{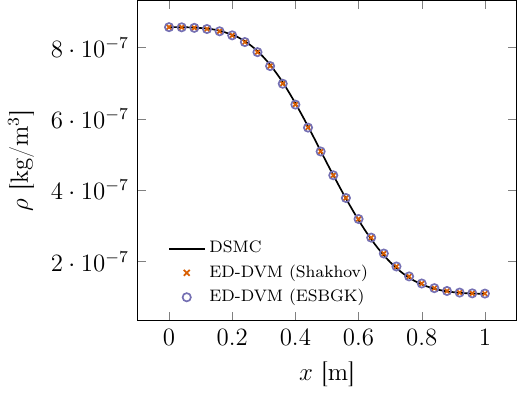}}} 
  \hfill
  \subfloat{\scalebox{0.7}{\includegraphics{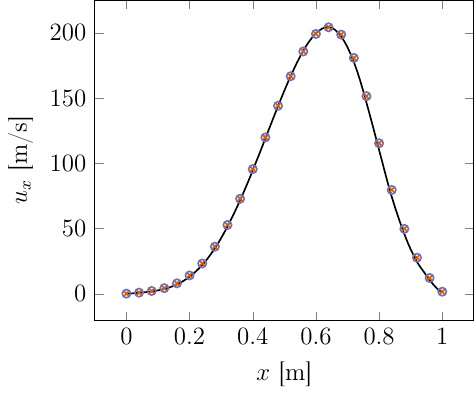}}} 
  \hfill
  \subfloat{\scalebox{0.7}{\includegraphics{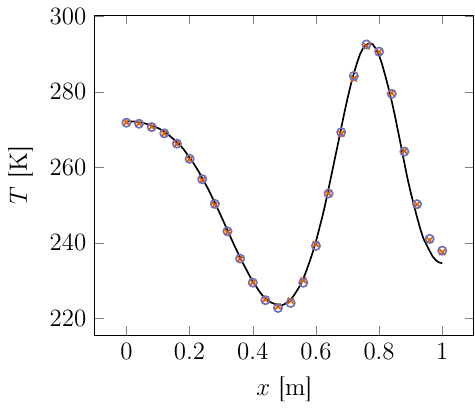}}} 
\end{adjustwidth}
  \caption{Sod shock tube at Kn $=1$: density, velocity and temperature profiles at $t=7\times 10^{-4}$ s.}
  \label{fig:sodrare}
\end{figure*}

The Variable Hard Sphere collision model was used for the DSMC reference simulations in order to match the definition of viscosity in the BGK equation. In the low density case, the DSMC reference results could be obtained with the same time step but in the near-continuum case, a 20 times smaller time step and adaptive subcells had to be used in order to resolve the mean collision time and the mean free path. Furthermore, the need of around $10^6$ particles per cell to reduce the stochastical noise induced by DSMC greatly exceeded the computational cost of ED-DVM on this unsteady flow that prevents any time averaging. Indeed, only 15 velocity points were needed in the dense case to match DSMC (Fig.~\ref{fig:soddense}), while 50 were enough to reach good accuracy in the rarefied regime (Fig.~\ref{fig:sodrare}). Besides, as it can be seen that the difference between DVM simulations using the ESBGK model and those using the Shakhov model are negligible, only the Shakhov model will be used in the following tests as it is the one with the simplest (and most efficient) implementation.

Table~\ref{tab:cputime} compares in more details the CPU time needed using the parallelized code PICLas, already heavily optimized for DSMC. The time required by ED-DVM is so short that most of the cost comes from the code initialization and the influence of the velocity grid size is negligible. Besides, as the acceptable level of noise in the DSMC results is a rather arbitrary choice, the number of simulation particles was here fixed so as to resolve the mean free path in the dense case. The times displayed for DSMC are therefore obtained using only 300 particles per cell in the left part of the domain (the noise-free DSMC simulations used for validation lasted about 8 hours in the dense case). This however resulted in a stochastical noise level leading to an error on the macroscopic variables of around 10\%. This clear advantage of the discrete velocity method over DSMC on this case would nonetheless disappear with increasing Knudsen number, Mach number or dimensionality, as an accurate deterministic discretization of the velocity space would become more complex.

\begin{table}
\caption{\label{tab:cputime}Time needed to solve the Sod shock tube problem on 10 cores of an AMD EPYC 7713 processor.}
\centering
\begin{tabular}{lcc}
Kn&DSMC&ED-DVM\\
\hline
0.01 & 4.59 s & 0.21 s\\
1 & 0.68 s & 0.29 s\\
\end{tabular}
\end{table}

\subsection{Force-driven Poiseuille flow}

In order to examine more in depth the spatio-temporal convergence of the ED-DVM, a force-driven Poiseuille flow was simulated at two different Knudsen numbers using various space and time resolutions. This test case consists in a one dimensional domain with a length of $L_y=1$ m, where an argon flow is accelerated by a constant force $F_x$ perpendicular to the domain, chosen here so that $\frac{\rho}{m}F_x=10^{-2}$ Pa/m, where $m$ is the molecular mass. This force is applied in two steps, adding half the force before and after every time step, resulting in a Strang splitting approach, as suggested in~\citet{wang2018comparative}. An equilibrium approximation is made here, so that each half force term can be simply added to the distribution function via:
\begin{equation}
  f_{new} = f_{old} + \frac{\Delta t}{2} \frac{F_x}{m} \frac{\partial f}{\partial v_x} \approx  f_{old} + \frac{\Delta t}{2} \frac{F_x c_x}{mRT} f^M.
  \label{eq:force}
\end{equation}
Up to a Knudsen number around 0.1, this approximation yields accurate results, but for more rarefied flows, the term $\frac{\partial f}{\partial v_x}$ is calculated using centered finite differences in the velocity space.
The boundaries at $y=0$ m and $y=1$ m are immobile and diffusive, with a wall temperature equal to the initial gas temperature $T_0=273$ K.

\begin{figure}
  \centering
  \subfloat{\includegraphics{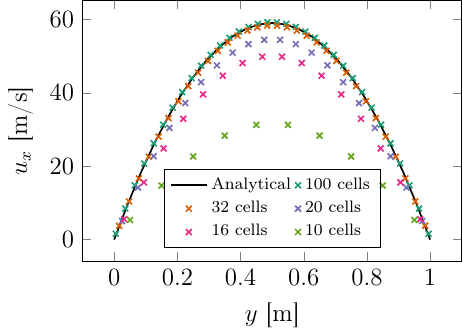}} 
  \caption{Dense Poiseuille flow: velocity profile obtained with ED-DVM using different resolutions.}
  \label{fig:poisdens}
\end{figure}

This Poiseuille flow is first simulated in near-continuum conditions, with $\rho = 8.58\times 10^{-5}$ kg/m$^3$ yielding Kn $\simeq 0.0013$, using a $5\times 5$ points Gauss-Hermite quadrature with reference temperature $T_{GH}=T_0$. The CFL number is fixed to 0.9 while the time step and spatial resolution are changed proportionally. The resulting velocity profiles are compared on Fig.~\ref{fig:poisdens} to the continuum analytical solution~\cite{myong2011full}:
\begin{equation}
  u_x = \frac{\rho F_x}{m} \frac{L_yy-y^2}{2\mu}.
\end{equation}
Very good agreement was obtained with the 100 cells mesh, where the relaxation factor is around $\nu \Delta t = 3$, with even acceptable accuracy with 32 cells and a relaxation factor as high as 9.5.

\begin{figure}
  \centering
  \subfloat{\includegraphics{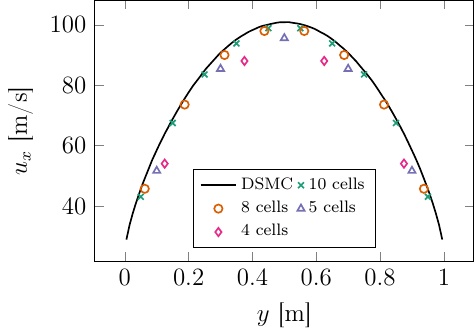}} 
  \caption{Rarefied Poiseuille flow: velocity profile obtained with ED-DVM using different resolutions.}
  \label{fig:poisrare}
\end{figure}

A more rarefied case at $\rho = 8.58\times 10^{-7}$ kg/m${-3}$ (Kn $\simeq 0.13$) is also simulated, and compared to a reference DSMC simulation on Fig.~\ref{fig:poisrare}. Regarding the velocity discretization, no substantial improvement in accuracy was obtained beyond $101\times 101$ equally spaced points between -3000 and 3000 m/s, so all ED-DVM simulations were performed with this quadrature. In this case, the low relaxation frequency combined to the CFL condition prevented us from investigating high relaxation factors, but with a CFL number of 0.9, a simple 10 cells mesh proved enough to accurately match the DSMC results. It should however be noted that the Shakhov model cannot retrieve the high order super-Burnett terms, which are necessary to produce the specific temperature and higher moments profiles at high Knudsen numbers for this flow \cite{xu2003superburnett}. Nevertheless, our method was sufficient in this case when focusing only on the velocity profile. As both DSMC and ED-DVM are explicit time-accurate methods, the steady state was reached at the same point in simulation time, but the DSMC solution is obtained by letting the simulation run after reaching that steady state, and averaging over time until statistical noise is reduced to reasonable levels.

\begin{figure}
  \centering
  \subfloat[]{\includegraphics{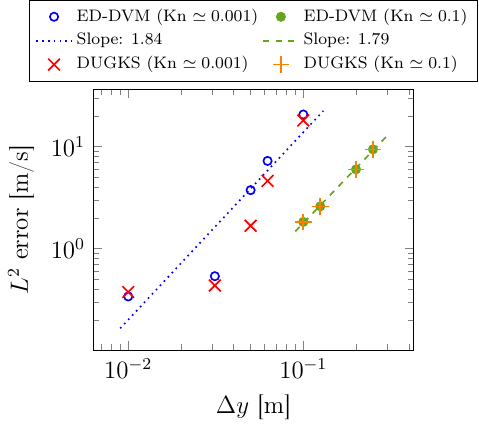}} 
  \caption{Poiseuille flow: error convergence of the velocity $u_x$ in the dense and rarefied cases.}
  \label{fig:poiserror}
\end{figure}

Furthermore, this analysis with varying spatio-temporal discretization allows the error convergence in the two cases to be shown on Fig.~\ref{fig:poiserror}. The same simulations were carried out with DUGKS. In the dense regime, better results were obtained with DUGKS at some discretization levels, but the overall convergence is similar to ED-DVM. In the rarefied case, the use of DUGKS lead to virtually the exact same results as with ED-DVM. Regardless of the rarefaction level, both methods display a convergence rate that is compatible with the theoretical second-order accuracy in space and time.

\subsection{Lid-driven cavity flow}

\begin{figure}
  \vspace {-1cm}
  \centering
  \begin{adjustwidth}{-9pt}{-9pt}
  \begin{tabular}{c c c}
      \multicolumn{3}{c}{\textbf{(a)} $\mathrm{Kn} = 0.1$, $\mathrm{Re} = 2.6$;\hspace{1cm} \textbf{(b)}  $\mathrm{Kn} = 2.6\times10^{-3}$, $\mathrm{Re} = 100$;} \\
      \multicolumn{3}{c}{ \textbf{(c)}  $\mathrm{Kn} = 2.6\times10^{-4}$, $\mathrm{Re} = 1000$;\hspace{1cm} \textbf{(d)}  $\mathrm{Kn} = 2.6\times10^{-5}$, $\mathrm{Re} = 10\ 000$} \\
      \begin{minipage}{0.4\textwidth}
          \centering
          \includegraphics[scale=0.22,trim={11.5cm 2cm 10cm 3.5cm},clip]{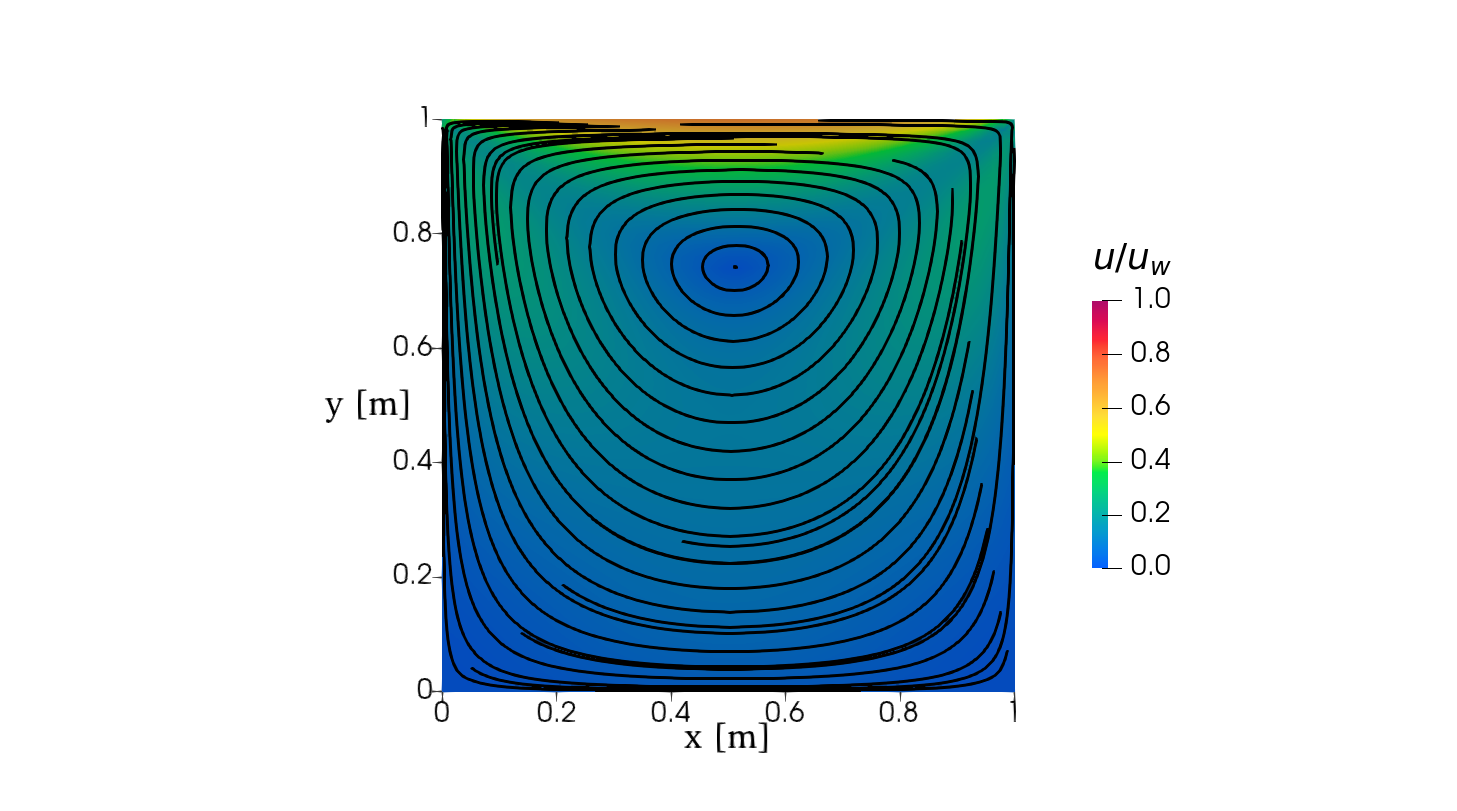}
      \end{minipage}
      & \textbf{(a)} & 
      \begin{minipage}{0.52\textwidth}
          \vspace{0.5cm}
          \hfill
          \scalebox{0.9}{\includegraphics{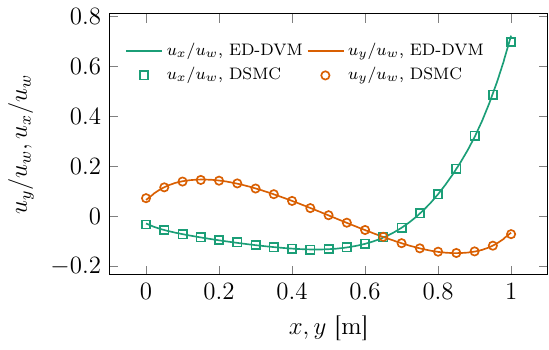}} 
      \end{minipage}
      \\

      \begin{minipage}{0.4\textwidth}
          \centering
          \includegraphics[scale=0.22,trim={11.5cm 2cm 10cm 3.5cm},clip]{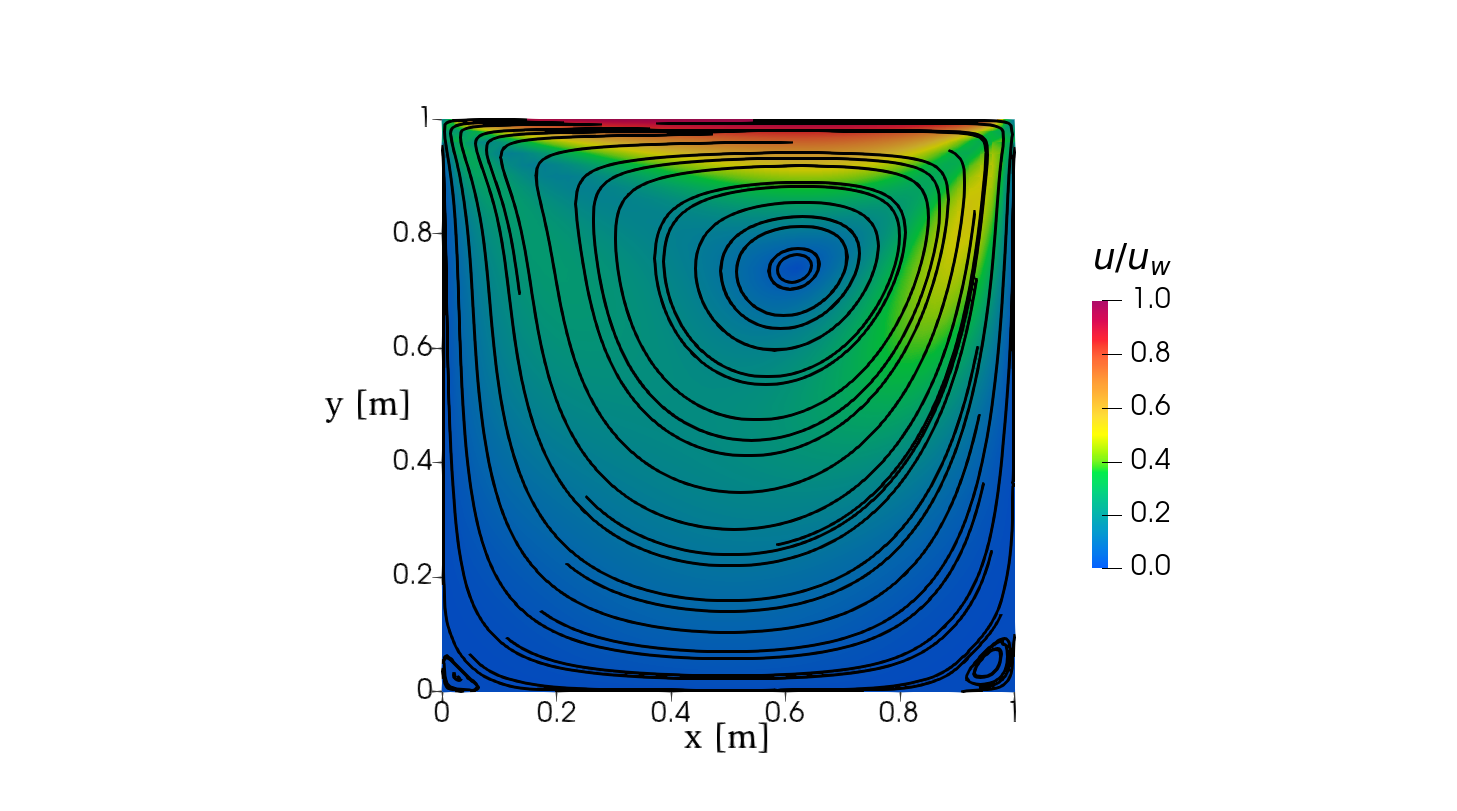}
      \end{minipage}
      & \textbf{(b)} & 
      \begin{minipage}{0.52\textwidth}
        \vspace{0.5cm}
          \hfill
          \scalebox{0.9}{\includegraphics{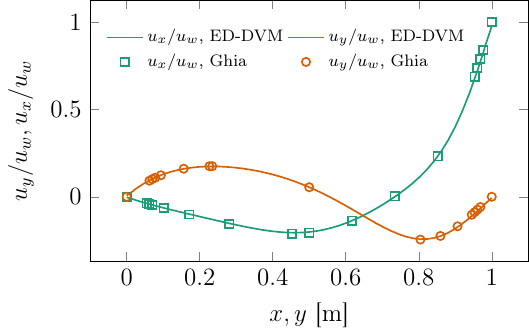}} 
      \end{minipage}
      \\

      \begin{minipage}{0.4\textwidth}
          \centering
          \includegraphics[scale=0.22,trim={11.5cm 2cm 10cm 3.5cm},clip]{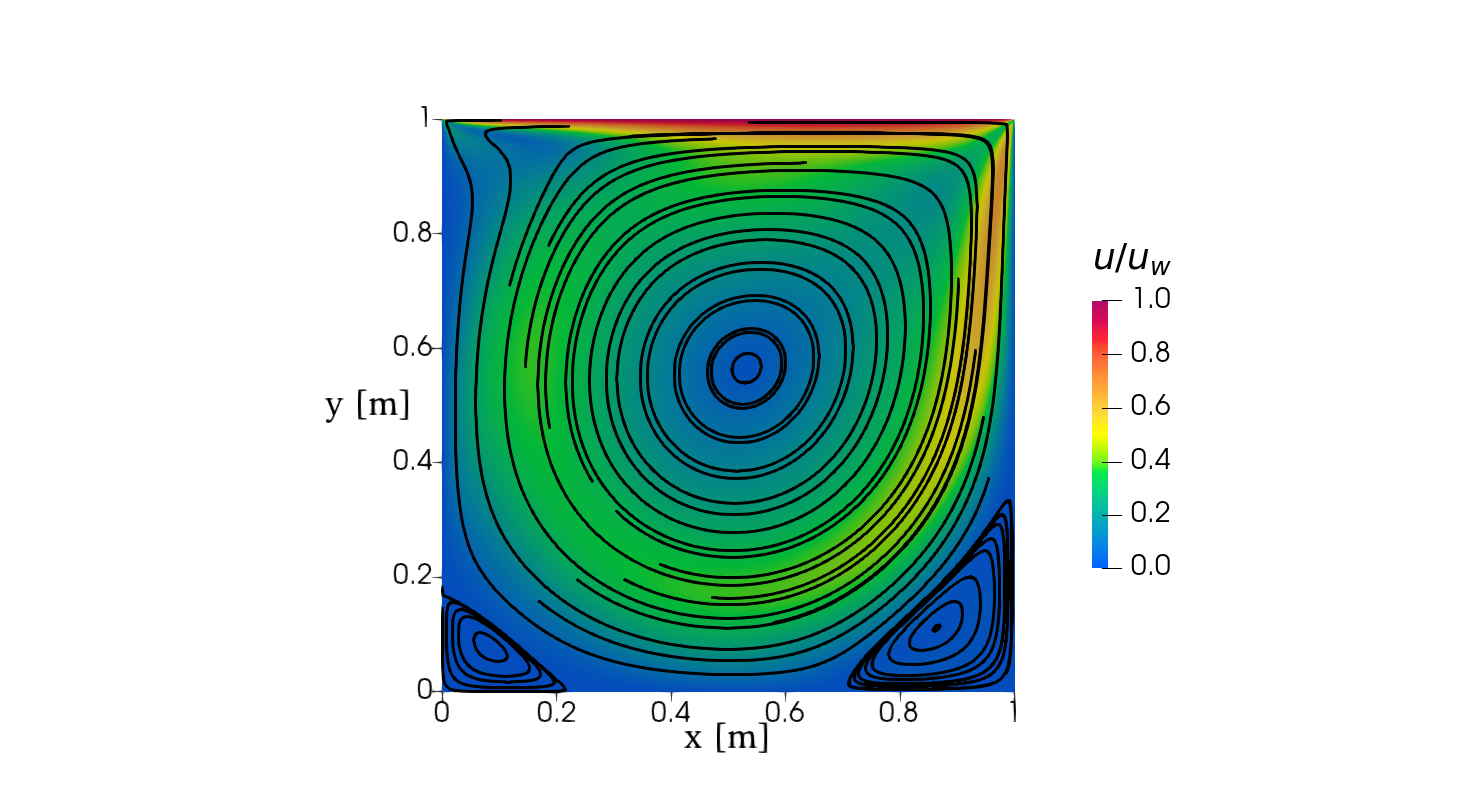}
      \end{minipage}
      & \textbf{(c)} & 
      \begin{minipage}{0.52\textwidth}
        \vspace{0.5cm}
          \hfill
          \scalebox{0.9}{\includegraphics{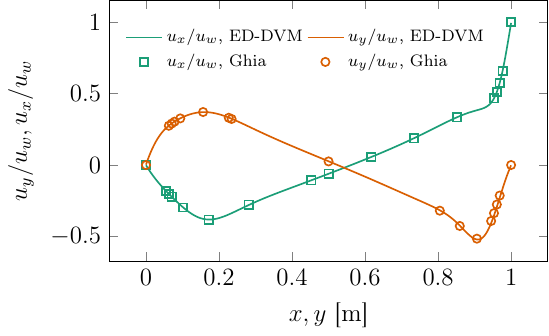}} 
      \end{minipage}
      \\

      \begin{minipage}{0.4\textwidth}
          \centering
          \includegraphics[scale=0.22,trim={11.5cm 2cm 10cm 3.5cm},clip]{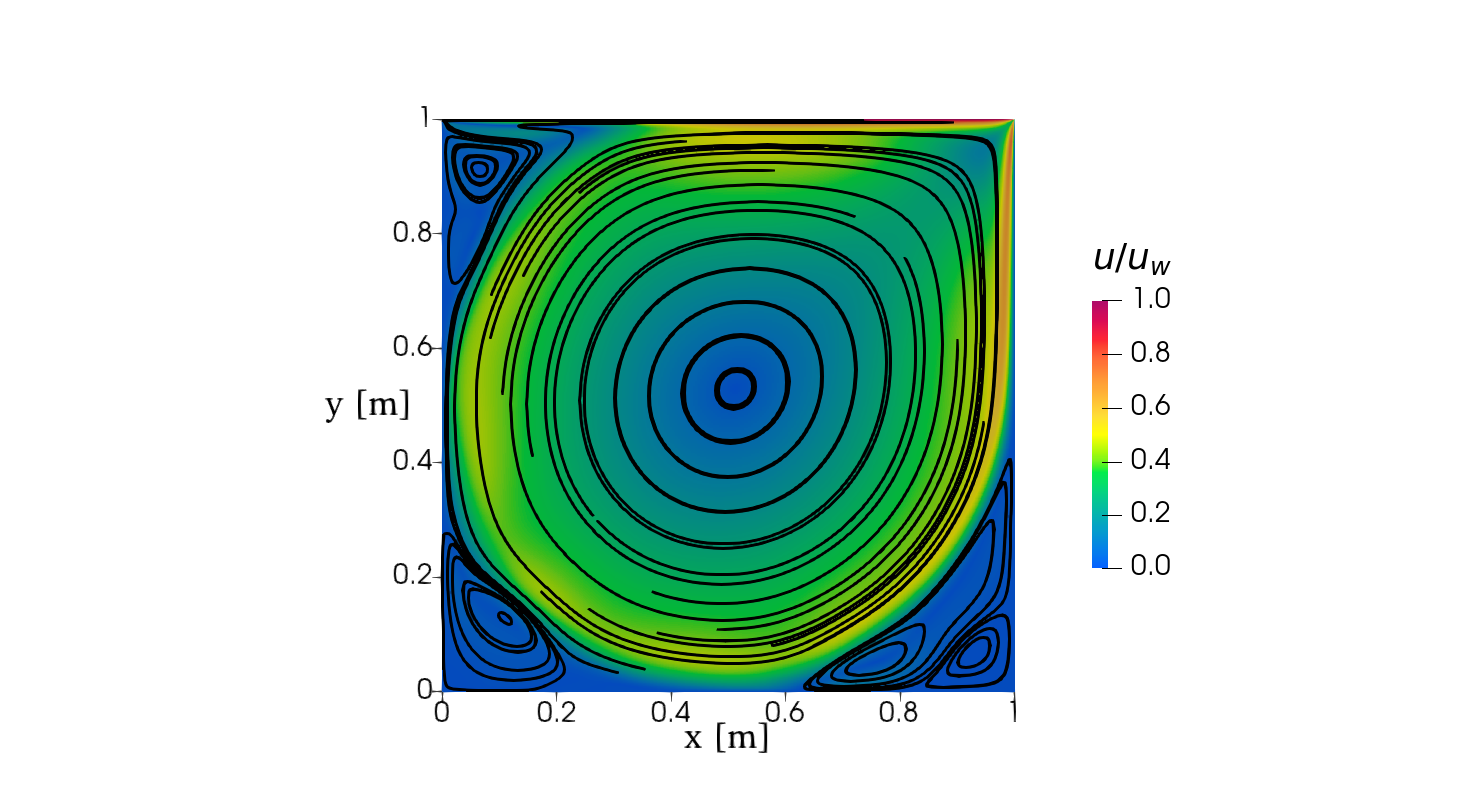}
          \caption{\centering Lid-driven cavity flow: velocity field lines and magnitude, simulated with ED-DVM.}
          \label{fig:lidstream}
      \end{minipage}
      & \textbf{(d)} & 
      \begin{minipage}{0.52\textwidth}
        \vspace{0.5cm}
          \hfill
          \scalebox{0.9}{\includegraphics{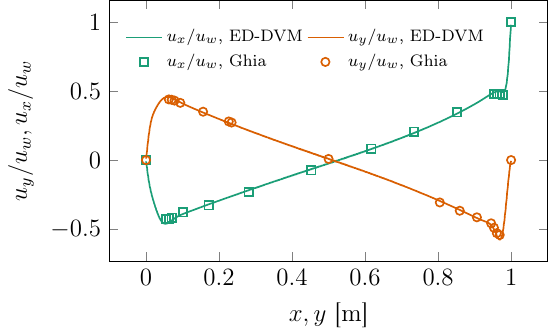}} 
          \caption{\centering Lid-driven cavity flow: $x$-velocity along a central vertical line (green) and $y$-velocity along a central horizontal line (orange).}
          \label{fig:lidvelolines}
      \end{minipage}
      \\
  \end{tabular}
  \end{adjustwidth}
\end{figure}

The lid-driven cavity flow was then simulated to observe the method's performance on a well-studied fully 2D case, with reference results available for a wide range of Reynolds numbers, obtained with Navier-Stokes solvers~\cite{ghia1982highre} or discrete velocity methods~\cite{guo2013discrete,fei2020unified}. Using a particle-based method to simulate this low Mach number flow would be much less efficient due to the particularly high number of particles needed to reach a reasonable signal-to-noise ratio. Indeed, this test case consists of a square cavity with diffusive boundaries where the flow of the initially motionless gas is only driven by the top wall moving at a constant slow velocity $u_w$ in its own plane. The initial and wall temperatures are set to $T_0=T_w=273$ K while we use the same lid velocity $u_w=50$ m/s in all cases, changing the argon density (and therefore the kinematic viscosity) to obtain the desired Reynolds number.

The mesh grid was adapted depending on the Reynolds number in order to resolve the arising secondary vortices, which can be seen on Fig.~\ref{fig:lidstream}. $50\times 50$ cells were used for Re = 100 (case (b)), $200\times 200$ for Re = 1000 (case (c)), while $800\times 800$ cells were necessary for accurate results for Re = 10 000 (case (d)), the two finer grids being non-uniform with higher resolution in vortex regions. A $5\times5$ Gauss-Hermite velocity grid was sufficient in all cases as the most rarefied case studied here only has a Knudsen number of $2.6\times10^{-3}$ and the low-speed nature of the flow limits the non-equilibrium effects. The CFL number of 0.9 was sufficient to match the reference results of~\citet{ghia1982highre} obtained with a Navier-Stokes solver.

A more rarefied simulation at Kn = 0.1 was also carried out (case (a)), using the $50\times 50$ uniform mesh, a $28\times 28$ Gauss-Hermite velocity grid and a CFL number of 0.9. As the Navier-Stokes equations do not give accurate results in this regime \cite{huang2012unified}, a highly resolved DSMC simulation was used as the reference in this case.

Comparisons to the references are shown on Fig.~\ref{fig:lidvelolines} where the values of the velocity components $u_x$ and $u_y$ are taken along a line perpendicular to the direction of the component and passing through the center of the cavity.
The ED-DVM was able to accurately capture the features of this flow even for the high Reynolds number of 10,000 where sharp variations in the velocity field arise.

\subsection{Flow around a cylinder}
\label{sec:cylinder}

\begin{figure}
  \centering
  \subfloat[]{\includegraphics[scale=0.25,trim={12cm 1cm 7cm 2cm},clip]{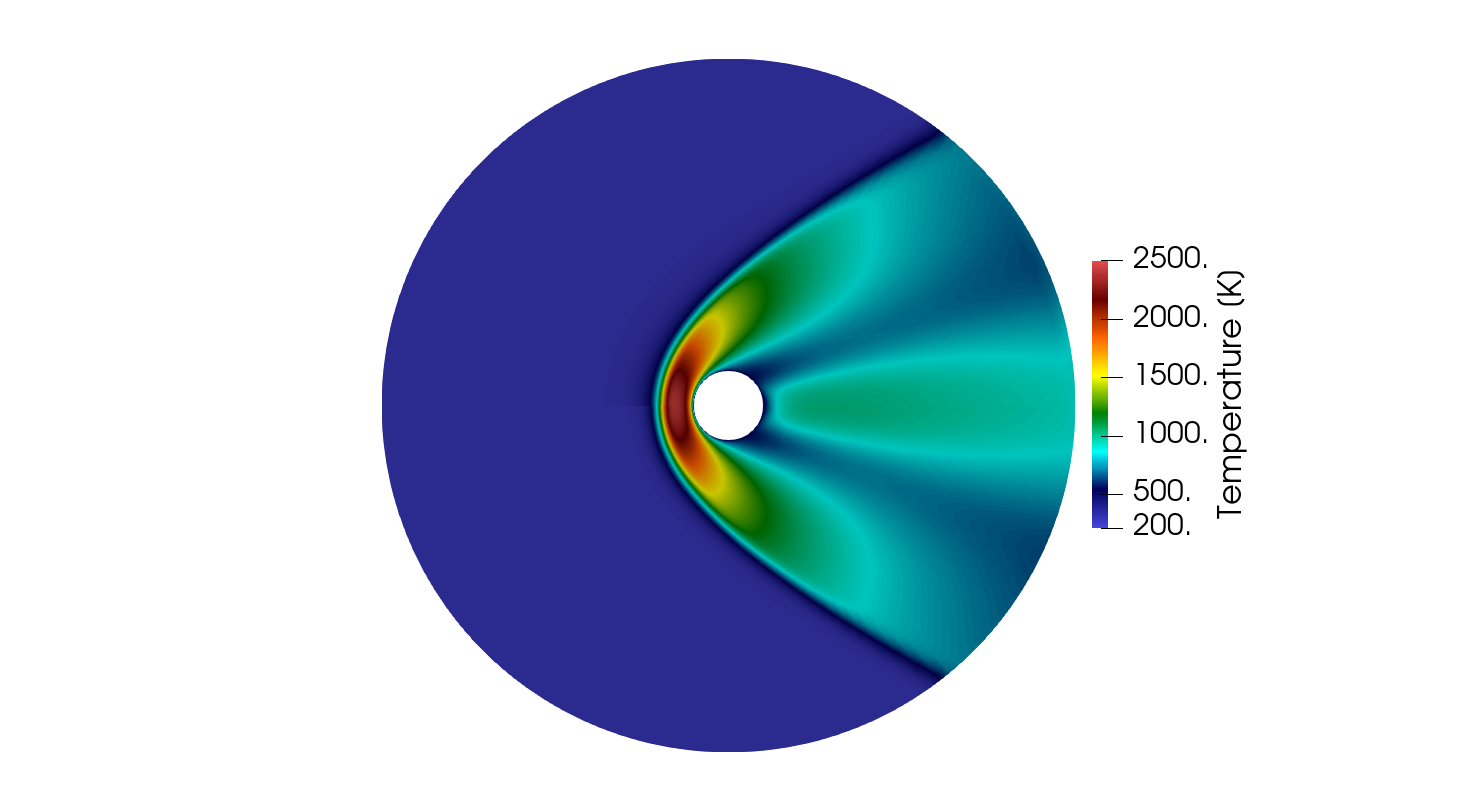}}
  \hfill
  \subfloat[]{\includegraphics{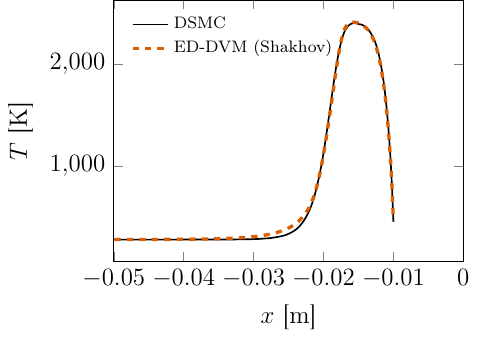}} 
  \caption{Temperature profile of the flow around a cylinder: (a) top half: ED-DVM, bottom half: DSMC, (b) along the horizontal line at $y=0$ m.}
  \label{fig:cyltemp}
\end{figure}

\begin{figure}
  \centering
  \subfloat{\scalebox{0.8}{\includegraphics{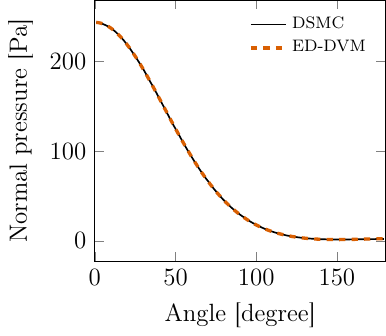}}} 
  \hfill
  \subfloat{\scalebox{0.8}{\includegraphics{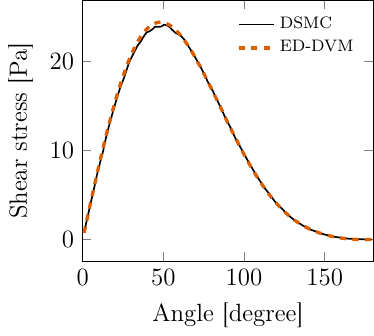}}} 
  \hfill
  \subfloat{\scalebox{0.8}{\includegraphics{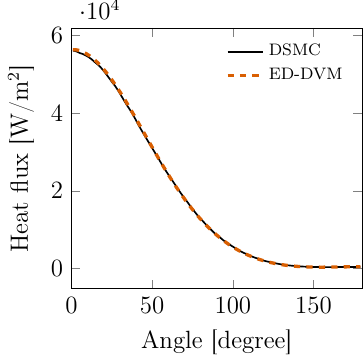}}} 
  \caption{Normal pressure, shear stress and heat flux profiles on the cylinder surface.}
  \label{fig:cylsurf}
\end{figure}

The last test case is a 2D hypersonic flow around an infinite cylinder. Argon flows past a cylinder of radius $r=1$ cm at a Mach number of 5, with a Knudsen number of $0.1$ in the inflow. The simulated domain is a half disk of radius $10r$, divided in 20 000 mesh cells with a higher resolution close to the obstacle to resolve the steep gradients induced by the shock. The velocity grid of $89 \times 89$ points between $-5055$ and $5055$ m/s was used, as it was shown by~\citet{zhu2016discrete} to yield accurate results with DUGKS. To better handle the hypersonic shock, the Shakhov model was again preferred and the~\citet{venkatakrishnan1995convergence} limiter was used. The CFL number was set to 0.9. The simulation is stopped at $t=10^{-3}$ s where steady state can largely be considered as reached because the flow temperature does not change in any point more than 0.01\% over the next 100 time steps. A comparison of the results to a reference DSMC simulation can be seen, for the temperature in the flow field on Figure \ref{fig:cyltemp}, and for the heat flux and stress exerted on the cylinder front surface on Figure \ref{fig:cylsurf} where the angle is taken from the horizontal.

\begin{figure}
  \centering
  \subfloat{\includegraphics{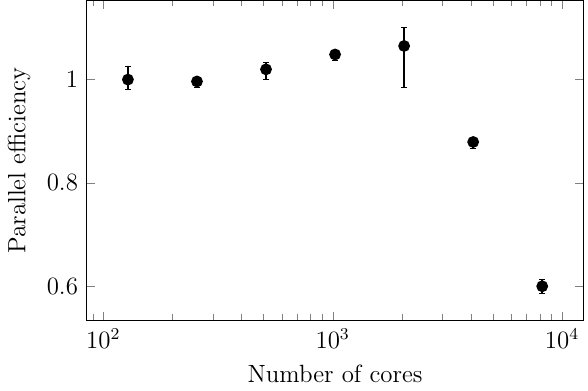}} 
  \caption{Strong scaling of the ED-DVM code on the Hawk supercomputer. The error bars indicate the extrema of the duration obtained over 5 simulation runs for each number of cores.}
  \label{fig:scaling}
\end{figure}

The scaling of our code was then evaluated on this case using up to 64 nodes of each 128 cores on the Hawk supercomputer of the High Performance Computing Center Stuttgart (HLRS). The parallel efficiency of ED-DVM, defined as $E=\frac{T_1}{pT_p}$ where $T_p$ is the running time of a simulation executed on $p$ processors, is displayed on Figure \ref{fig:scaling}. Using a similar MPI framework as already used for the electromagnetic field solver of PICLas~\cite{ortwein2015parallel}, the efficiency remains ideal until the number of cores approaches the number of mesh cells and work load becomes unbalanced. Some superlinear scaling with an efficiency above 1 can even be observed, probably due to a suboptimal usage of the cache memory that slightly lowers the performance when simulating on a small number of cores.

\begin{figure}
  \centering
  \subfloat{\scalebox{0.8}{\includegraphics{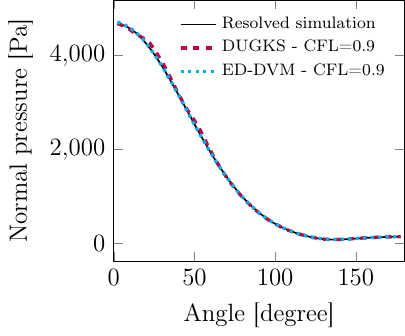}}} 
  \hfill
  \subfloat{\scalebox{0.8}{\includegraphics{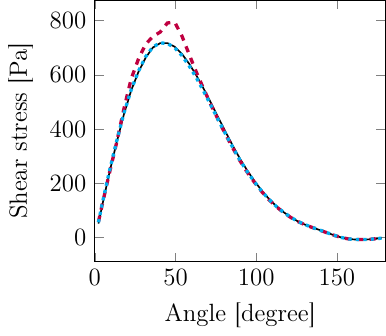}}} 
  \hfill
  \subfloat{\scalebox{0.8}{\includegraphics{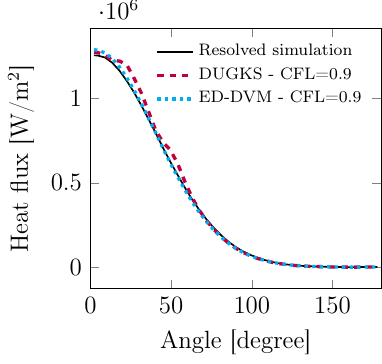}}} 
  \caption{Normal pressure, shear stress and heat flux profiles on the cylinder surface.}
  \label{fig:cylsurfcoarse}
\end{figure}

Furthermore, as this flow contains non-equilibrium regions arising not from a simple rarefaction but from the supersonic shock, it was simulated again in conditions set to show the difference of stability between ED-DVM and DUGKS. The ESBGK model is preferred here for its positivity, so that no instabilities are introduced via the target distribution. The density of the inflow was increased, giving a Knudsen number of Kn $= 5\times 10^{-4}$, while a coarser mesh of 2000 cells was used. Keeping the other simulation parameters from the previous case, notably the CFL number of 0.9, this resulted in a high relaxation factor of around 5 close to the cylinder. The results on the cylinder surface are compared on Figure~\ref{fig:cylsurfcoarse}, displaying more accurate results for ED-DVM than for DUGKS. The reference is a DVM simulation with a reduced time step, corresponding to a CFL number of 0.1 (both ED-DVM and DUGKS converged to this same result with this small time step). The better behaviour of ED-DVM in such an under-resolved simulation was expected, as this is the type of dense non-equilibrium cases where DUGKS can lead to negative values in its solution.

\section{Conclusion}

A second-order asymptotic preserving BGK solver was presented and evaluated on various test cases and in different regimes.
Although the performances are similar to pre-existing DVM schemes such as DUGKS, the main advantage of ED-DVM is that all operations corresponding to the relaxation of the distribution function ensure the positivity of every term. This feature, combined to the asymptotic preserving and second-order properties reached without introducing implicit steps in the final algorithm, allows for a straightforward correspondence between high accuracy deterministic and stochastic schemes, now both available in the open-source simulation framework PICLas.

While the ED-DVM has the usual advantages inherent to DVMs compared to particle solvers when it comes to low-velocity near-continuum flows, its particle adaptation presented in~\citet{pfeiffer2022exponential} can become more efficient in rarefied and high-speed regimes. In order to fully benefit from the efficiency of the exponential differencing approach, a coupling between the two solvers, either in velocity or physical space, would be valuable as future development. Besides, the ED-DVM could be extended to simulate molecular gases or more costly cases with better optimized velocity grids. PICLas' new DVM module, validated here on one and two-dimensional test cases but already allowing for full 3D simulations, could then be applied to the study of more complex flows~\cite{baranger2014locally,zhao2024interaction}.

\section*{Acknowledgments}
  This project has received funding from the European Research Council (ERC) under the European Union's Horizon 2020 research and innovation programme (grant agreement No. 899981 MEDUSA).

  The authors also thank the High Performance Computing Center Stuttgart (HLRS) for granting the computational time that has allowed the execution of the presented simulations.

\appendix
\section{Spatio-temporal convergence analysis}
\label{apx:order}

If the flux term that arises from exponential differencing (see Eq. \ref{eq:exactEI}) could be computed exactly, the only source of error in the time-discretized scheme would be the linear approximation of the target (Eq. \ref{eq:linearapprox}). Without space discretization, a step of the ED-DVM with exact fluxes, when the time step goes to zero, can therefore be written as:

\begin{equation}
    \gamma \widetilde f_{n+1} = \gamma \hat f_n - \mathcal F_n + O(\Delta t^3), \qquad \gamma=\frac{1-e^{-\nu\Delta t}}{\nu\Delta t}
\end{equation}

\begin{equation}
\text{where}\qquad\mathcal F_n = \int_{0}^{\Delta t} e^{\nu(s-\Delta t)}\, \mathbf v \cdot \frac{\partial f}{\partial \mathbf x}(\mathbf x, \mathbf v, t^n+s)\,ds
\end{equation}

The procedure described in Section \ref{sec:flux} can then be applied to the spatially discretized flux term:
\begin{eqnarray}
\overline{\mathcal F_{j,n}} &=& \frac{1}{|V_j|}\int_{\mathbf x \in V_j}\int_{0}^{\Delta t} e^{\nu(s-\Delta t)}\, \mathbf v \cdot \frac{\partial f}{\partial \mathbf x}(\mathbf x, \mathbf v, t^n+s)\,ds\,d\mathbf x \nonumber\\
&=& \frac{1}{|V_j|}\int_{0}^{\Delta t} e^{\nu(s-\Delta t)}\, \int_{\mathbf x \in \partial V_j}(\mathbf v \cdot\mathbf n) f(\mathbf x, \mathbf v, t^n+s)\,d\mathbf x\,ds \nonumber\\
&=& \sum_{S_k \in \partial V_j} \left(\frac{|S_k|}{|V_j|}\Delta t \gamma (\mathbf v \cdot\mathbf n_k) \left[f(\mathbf x_k, \mathbf v, t^n+\frac{\Delta t}{2}) + O(\Delta x^2) + O(\Delta t^3)\right]\right) \nonumber\\
&=& \gamma\Delta t  F_{j,n+1/2} + O(\Delta x^2) + O(\Delta t^3)
\end{eqnarray}

As the construction of $F_{j,n+1/2}$ uses only exact exponential integration along characteristic lines and a second order gradient calculation, the approximation made in the reconstruction of the distribution at the interface does not reduce the order of accuracy:
\begin{eqnarray}
\widetilde f^r(\mathbf x_k, \mathbf v, t^n+\frac{\Delta t}{2}) &=& \hat f^r(\mathbf  x_k - \mathbf v \frac{\Delta t}{2}, \mathbf v, t^n) \nonumber\\
&=& \hat f^r(\mathbf x_j, \mathbf v, t^n) + (\mathbf x_k - \mathbf{x_j} - \mathbf v \frac{\Delta t}{2})\cdot\nabla\hat f^r(\mathbf x_j, \mathbf v, t^n) + O(\Delta x^2)
\end{eqnarray}

With all approximations, the final ED-DVM scheme is therefore second-order accurate in time and space:
\begin{equation}
     \frac{\widetilde f_{j,n+1} - \hat f_{j,n}}{\Delta t} = - F_{j,n+1/2} + O(\Delta x^2) + O(\Delta t^2).
\end{equation}

\section{Asymptotic analysis}
\label{apx:AP}
In the free molecular flow limit ($\nu \rightarrow 0$), using $e^x \sim 1 + x$ yields $\hat{f}=\widetilde{f}=\hat{f^r}=\widetilde{f^r}=f$. The analysis of \ref{apx:order} can therefore be repeated with many simplifications, proving that the ED-DVM becomes a simple second-order finite volume scheme solving an advection equation:
\begin{equation}
  \frac{f_{j,n+1} - f_{j,n}}{\Delta t} = - F_{j,n+1/2} + O(\Delta x^2) + O(\Delta t^2).
\end{equation}

In the continuum limit, as $\gamma \underset{\nu \to \infty}{\longrightarrow} 1$, we get $\widetilde f \underset{\nu \to \infty}{\longrightarrow} f$ but $\hat f\underset{\nu \to \infty}{\longrightarrow} f^t$, making the scheme equivalent to

\begin{equation}
  \label{eq:fcontinuum}
  \frac{f_{j,n+1} - f^t_{j,n}}{\Delta t} = - F_{j,n+1/2} + O(\Delta x^2) + O(\Delta t^2).
\end{equation}
Since the conservative moments are the same for $f$ and $f^t$, multiplying Eq. \eqref{eq:fcontinuum} by $\bm{\psi}(\mathbf v)$ and integrating over the velocity space gives:
\begin{equation}
  \label{eq:NSsolver}
  \frac{\mathbf W_{j,n+1} - \mathbf W_{j,n}}{\Delta t} = -\bm{\Phi}_{j,n+1/2} + O(\Delta x^2) + O(\Delta t^2)
\end{equation}
where
\begin{equation}
  \bm{\Phi}_{j,n+1/2} = \int_{\mathbb R^3} \bm{\psi}(\mathbf v) F_{j,n+1/2}(\mathbf v) d\mathbf v.
\end{equation}
To prove the asymptotic preserving property, it remains only to show that $\bm{\Phi}_{j,n+1/2}$ is a consistent flux for the Navier-Stokes equations.

The approach of \citet{guo2013discrete} can then be followed and the distribution function is expressed by its Chapman-Enskog expansion:
\begin{equation}
  \label{eq:CEexp}
  f(t) = f^t(t) -\frac{1}{\nu} D_t f^t(t) + O(\nu^{-2})
\end{equation}
where $D_t= (\partial_t + \mathbf v \cdot \nabla)$.
The flux construction is carried out again like in Section \ref{sec:flux}, now in the asymptotic limit $\nu\rightarrow \infty$.

Rewriting Eq.~\eqref{eq:ftilde^r} as
\begin{equation}
  f = \gamma^r \widetilde{f}^r + (1-\gamma^r)f^t
\end{equation}
and using Eqs.~\eqref{eq:halfstep} and~\eqref{eq:fhat^r}, the reconstruction of the distribution at a cell boundary becomes, with $h=\Delta t/2$ and omitting the dependency of $f$ on $\mathbf v$ for more clarity:
\begin{eqnarray}
  f(\mathbf x_b, t+h) & = & \gamma^r \hat{f}^r(\mathbf x_b - \mathbf v h, t) + (1-\gamma^r)f^t(\mathbf x_b, t+h) \nonumber \\
  & = & e^{-\nu h} f(\mathbf x_b - \mathbf v h, t) + (\gamma^r - e^{-\nu h})f^t(\mathbf x_b - \mathbf v h, t) + (1-\gamma^r)f^t(\mathbf x_b, t+h) \nonumber \\
  &=& e^{-\nu h} \left[f(\mathbf x_b, t) - h\mathbf v \cdot \nabla f(\mathbf x_b, t)\right] + (\gamma^r - e^{-\nu h})\left[f^t(\mathbf x_b, t) - h\mathbf v \cdot \nabla f^t(\mathbf x_b, t)\right] \nonumber \\
  &&  + (1-\gamma^r)\left[f^t(\mathbf x_b, t) + h\partial_t f^t(\mathbf x_b, t)\right] + O(h^2)
\end{eqnarray}
Using Eq.~\eqref{eq:CEexp} then yields:
\begin{eqnarray}
  f(\mathbf x_b, t+h) & = & e^{-\nu h} \left[f^t(\mathbf x_b, t) - \frac{1}{\nu} D_tf^t(\mathbf x_b, t)\right] + (\gamma^r-e^{-\nu h}) f^t(\mathbf x_b, t) \nonumber \\
  && - h \mathbf v \cdot \left(e^{-\nu h} \nabla f^t(\mathbf x_b, t) + (\gamma^r- e^{-\nu h})\nabla f^t(\mathbf x_b, t) \right) \nonumber \\
  && +(1-\gamma^r)\left[f^t(\mathbf x_b, t) + h\partial_t f^t(\mathbf x_b, t)\right] + O(h^2) + O(\nu^{-2})\nonumber \\
  & = & f^t(\mathbf x_b, t) - \frac{1}{\nu} D_t f^t(\mathbf x_b, t)+ h\partial_t f^t(\mathbf x_b, t) + O(h^2) + O(\nu^{-2})
\end{eqnarray}
which is in fact an expansion in time of a Chapman-Enskog distribution that has been shown to correspond to the Navier-Stokes model, when replacing $f^t$ either with the ESBGK~\cite{andries2000gaussianbgk} or the Shakhov~\cite{chen2015comparison} target distribution.

$\bm{\Phi}_{j,n+1/2}$ is then a second-order finite volume flux of the moments of this distribution function, it is therefore consistent with the Navier-Stokes equations. Hence, in this continuum limit, the ED-DVM becomes a Navier-Stokes solver described by~\eqref{eq:NSsolver}, retaining second-order accuracy in time and space.

\section{Positivity of the scheme}
\label{apx:positive}

To determine under which conditions the ED-DVM is a positive scheme, the positivity of all steps of the algorithm summarized in Section~\ref{sec:algo} is analyzed here.

Starting with a positive function $\widetilde{f}_n$ ($f_0$ in the first time step), step 1 is clearly positive as long as the target distribution remains positive. This is always the case with the ESBGK model, while the Shakhov model can lead to negative values, usually without any consequence for small deviations from the equilibrium~\cite{shakhov1968generalization}.

Steps 2 and 7 are simple linear combinations of distribution functions using positive coefficients (mentioned earlier as pre-factors). This is clear for equation~\eqref{eq:finallincom}, while it is proven for equation~\eqref{eq:ftildetofhat^r} by writing
\begin{equation}
  0<\frac{\gamma}{\gamma^r}=\frac{1-e^{-\nu\Delta t}}{2(1-e^{-\frac{\nu\Delta t}{2}})} = \frac{1+e^{-\frac{\nu\Delta t}{2}}}{2}< 1\quad \text{for}\quad \nu\Delta t > 0.
\end{equation}
During the first time step, steps 2 and 7 instead use equations~\eqref{eq:fhat} and~\eqref{eq:fhat^r}. In this case, using the convexity of the exponential function on $\left]0;+\infty\right[$, we write $e^{\nu\Delta t} - 1 > \nu \Delta t$, hence
\begin{equation}
  A = \frac{1}{\nu \Delta t} - \frac{e^{-\nu\Delta t}}{1-e^{-\nu\Delta t}} = \frac{e^{\nu\Delta t} - 1 - \nu \Delta t}{\nu \Delta t(e^{\nu\Delta t} - 1)}>0.
\end{equation}
This proves that~\eqref{eq:fhat} is also a positive linear combination of distributions, and it can be done similarly for~\eqref{eq:fhat^r}. The positivity of the solution is therefore always preserved during steps 2 and 7.

Step 8, where the solution is updated by taking the flux into account, only preserves positivity if the construction of the flux (steps 3 to 6) ensures it. Due to the complexity of a full study of the stability properties of the reconstruction using gradient limiters, both for ED-DVM or DUGKS, we restrict ourselves here to the case where the limiter would completely nullify the gradients. This would obviously remove the second-order property of the scheme, but it allows to easily show that, with the flux term as well, the positivity of the overall scheme is easier to achieve with ED-DVM than with DUGKS.

Looking at equation~\eqref{eq:update}, the solution of ED-DVM remains positive in cell $j$ if\\ ${\Delta t F_{j,n+1/2} \leq \hat{f}_{j,n}}$. As the incoming contribution from neighbouring cells to the flux term $F_{j,n+1/2}$ is always negative (due to our upwind framework with outward-pointing normal vectors), only the positive outgoing flux, calculated from the cell-local distribution, is considered. Since the reconstruction gradients are artificially set to zero, this positive flux term can be written as
\begin{eqnarray}
F^{+}_{j,n+1/2}&=&\sum_{S_k  \in \partial V_j} \frac{|S_k|}{|V_j|}(\mathbf n_k \cdot \mathbf v) \left(\gamma^r\widetilde{f}^r(\mathbf x_k, \mathbf v, t_n+\frac{\Delta t}{2})+(1-\gamma^r)f^t_{n+1/2}(\mathbf v)\right) \nonumber \\
&=&\sum_{S_k  \in \partial V_j} \frac{|S_k|}{|V_j|}(\mathbf n_k \cdot \mathbf v) \left(\gamma^r\hat{f}^r(\mathbf x_j, \mathbf v, t_n)+(1-\gamma^r)f^t_{n}(\mathbf v)\right) \nonumber \\
&=&\sum_{S_k  \in \partial V_j} \frac{|S_k|}{|V_j|}(\mathbf n_k \cdot \mathbf v) \left(e^{-\nu\frac{\Delta t}{2}} f(\mathbf x_j, \mathbf v, t_n)+(1-e^{-\nu\frac{\Delta t}{2}})f^t_{n}(\mathbf v)\right).
\end{eqnarray}
Therefore, we have
\begin{equation}\Delta t F_{j,n+1/2}\leq C(\Delta t) \left(e^{-\nu\frac{\Delta t}{2}} f(\mathbf x_j, \mathbf v, t_n)+(1-e^{-\nu\frac{\Delta t}{2}})f^t_{n}(\mathbf v)\right)
\end{equation}
where $C(\Delta t)$ is the CFL number. By comparing this upper bound to $\hat{f}_{j,n}$, a condition for the positivity of equation~\eqref{eq:update}, with any positive $f$ and $f^t$, can then be derived:
\begin{equation}
  C(\Delta t)\leq C_{\textrm{ED-DVM}}(\Delta t)=\frac{\nu \Delta t e^{-\nu\frac{\Delta t}{2}}}{1-e^{-\nu\Delta t}}.
\end{equation}
A CFL-like condition appears where the CFL number is restricted further below 1. However, this is not a strict condition as $C(\Delta t)> C_{\textrm{ED-DVM}}(\Delta t)$ does not usually lead to any negative value in the discretized velocity domain when $f$ is close to $f^t$. Taking in account the reconstruction gradients with a proper limiter could also reduce the chance of negative values as the distribution function is further relaxed towards the target. In practice, any CFL number below 1 lead to stable results in our test cases. Besides, conducting the same analysis with DUGKS gives a similar condition that can be shown to be even more restrictive:
\begin{equation}
  C(\Delta t)\leq C_{\textrm{DUGKS}}(\Delta t)=\frac{(2-\nu\Delta t)(2+\nu\frac{\Delta t}{2})}{4-\nu\Delta t}.
\end{equation}
This condition for DUGKS is only valid for $\nu\Delta t<2$, beyond which the positivity of the DUGKS flux step cannot be ensured in this manner.

\bibliographystyle{elsarticle-num-names}
\bibliography{biblio}

\end{document}